\documentclass[10pt]{article} 
\usepackage{float}
\usepackage{psfrag}
\usepackage{a4}
\usepackage{amsmath}
\usepackage[final]{graphicx}    
\usepackage{caption2}           
\usepackage{float}              
\usepackage{latexsym}            
\usepackage{exscale}
\include{mydefs}          
\usepackage{amsfonts}   

\newcommand{\RR}{\mathbb{R}}                                  

\begin{document}
\title{Graph Theory and Networks in Biology}
\author{Oliver Mason and Mark Verwoerd}
\maketitle
\begin{abstract}
In this paper, we present a survey of the use of graph theoretical
techniques in Biology.  In particular, we discuss recent work on
identifying and modelling the structure of bio-molecular networks,
as well as the application of centrality measures to interaction
networks and research on the hierarchical structure of such
networks and network motifs. Work on the link between structural
network properties and dynamics is also described, with emphasis
on synchronization and disease propagation.
\end{abstract}
\section{Introduction and Motivation}\label{sec:intro}
The theory of complex networks plays an important role in a wide
variety of disciplines, ranging from communications and power
systems engineering to molecular and population biology
\cite{AlbBar02, BarOlt04, New03, DorMen02, AlmArk03, AlbJeoBar99,
Bra03, Alo03}. While the focus of this article is on biological
applications of the theory of graphs and networks, there are also
several other domains in which networks play a crucial role. For
instance, the Internet and the World Wide Web (WWW) have grown at
a remarkable rate, both in size and importance, in recent years,
leading to a pressing need both for systematic methods of
analysing such networks as well as a thorough understanding of
their properties. Moreover, in sociology and ecology, increasing
amounts of data on food-webs and the structure of human social
networks are becoming available. Given the critical role that
these networks play in many key questions relating to the
environment and public health, it is hardly surprising that
researchers in ecology and epidemiology have focussed attention on
network analysis in recent years.  In particular, the complex
interplay between the structure of social networks and the spread
of disease is a topic of critical importance.  The threats to
human health posed by new infectious diseases such as the SARS
virus and the Asian bird flu \cite{WanRua04, Meyetal05}, coupled
with modern travel patterns, underline the vital nature of this
issue.

On a more theoretical level, several recent studies have indicated
that networks from a broad range of application areas share common
structural properties. Furthermore, a number of the properties
observed in such real world networks are incompatible with those
of the random graphs which had been traditionally employed as
modelling tools for complex networks \cite{AlbBar02, New03}. The
latter observation naturally poses the challenge of devising more
accurate models for the topologies observed in biological and
technological networks, while the former further motivates the
development of analysis tools for complex networks. The common
structural properties shared by diverse networks offers the hope
that such tools may prove useful for applications in a wide
variety of disciplines.  Within the fields of Biology and
Medicine, applications include the identification of drug targets,
determining the role of proteins or genes of unknown function
\cite{JeoOltBar03, SamLia03}, the design of effective containment
strategies for infectious diseases \cite{Eubetal04}, and the early
diagnosis of neurological disorders through detecting abnormal
patterns of neural synchronization in specific brain regions
\cite{SchGro05}.

Recent advances in the development of high-throughput techniques
in molecular biology have led to an unprecedented amount of data
becoming available on key cellular networks in a variety of simple
organisms \cite{Itoetal01, Cosetal00}.  Broadly speaking, three
classes of bio-molecular networks have attracted most attention to
date: metabolic networks of biochemical reactions between
metabolic substrates; protein interaction networks consisting of
the physical interactions between an organism's proteins; and the
transcriptional regulatory networks which describe the regulatory
interactions between different genes.  At the time of writing, the
central metabolic networks of numerous bacterial organisms have
been mapped \cite{Ravetal02}. Also, large scale data sets are
available on the structure of the protein interaction networks of
{\it S. cerevisiae} \cite{Itoetal01, Uetzetal00}, {\it H. pylori}
\cite{Raietal01}, {\it D. melanogaster} \cite{Gioetal03} and {\it
C. elegans} \cite{Lietal04, Cosetal00}, and the transcriptional
regulatory networks of {\it E. coli} and {\it S. cerevisiae} have
been extensively studied \cite{Ihmetal02, Shenetal02}.  The large
amount of data now available on these networks provides the
network research community with both opportunities and challenges.

On the one hand, it is now possible to investigate the structural
properties of networks in living cells, to identify their key
properties and to hopefully shed light on how such properties may
have evolved biologically. A major motivation for the study of
biological networks is the need for tailored analysis methods
which can extract meaningful biological information from the data
becoming available through the efforts of experimentalists.  This
is all the more pertinent given that the network structures
emerging from the results of high-throughput techniques are too
complex to analyse in a non-systematic fashion. A knowledge of the
topologies of biological networks, and of their impact on
biological processes, is needed if we are to fully understand, and
develop more sophisticated treatment strategies for, complex
diseases such as cancer \cite{VogLanLev00}.  Also, recent work
suggesting connections between abnormal neural synchronization and
neurological disorders such as {\it Parkinson's disease} and {\it
Schizophrenia} \cite{SchGro05} provides strong motivation for
studying how network structure influences the emergence of
synchronization between interconnected dynamical systems.

The mathematical discipline which underpins the study of complex
networks in Biology and elsewhere, and on which the techniques
discussed throughout this article are based, is {\it graph theory}
\cite{Die00}.  Alongside the potential benefits of applying graph
theoretical methods in molecular biology, it should be emphasized
that the complexity of the networks encountered in cellular
biology and the mechanisms behind their emergence presents the
network researcher with numerous challenges and difficulties.  The
inherent variability in biological data, the high likelihood of
data inaccuracy \cite{Meretal02} and the need to incorporate
dynamics and network topology in the analysis of biological
systems are just a sample of the obstacles to be overcome if we
are to successfully understand the fundamental networks involved
in the operation of living cells.  Another important issue, which
we shall discuss at various points throughout the article, is that
the structure of biological and social networks is often inferred
from sampled subnetworks.  The precise impact of sampling on the
results and techniques published in the recent past needs to be
understood if these are to be reliably applied to real biological
data.

Motivated by the considerations outlined above, a substantial
literature dedicated to the analysis of biological networks has
emerged in the last few years, and some significant progress has
been made on identifying and interpreting the structure of such
networks. Our primary goal in the present article is to provide as
broad a survey as possible of the major advances made in this
field in the recent past, highlighting what has been achieved as
well as some of the most significant open issues that need to be
addressed.  The material discussed in the article can be divided
naturally into two strands, and this is reflected in the
organisation of the document.  The first part of the article will
primarily be concerned with the properties and analysis of
cellular networks such as protein interaction networks and
transcriptional regulatory networks.  In the second part, we turn
our attention to two important applications of Graph Theory in
Biology: the phenomenon of synchronisation and its role in
neurological disorders, and the interaction between network
structure and epidemic dynamics.

In the interests of clarity, we shall now give a brief outline of
the main topics covered throughout the rest of the paper. In
Section 2, we shall fix the principal notations used throughout
the paper, and briefly review the main mathematical and graph
theoretical concepts that are required in the remainder of the
article.  As mentioned above, the body of the article is divided
into two parts.  The first part consists of Sections 3, 4 and 5
and the second part of Sections 6 and 7.  At the end of each major
section, a brief summary of the main points covered in that
section is given.

In Section 3, we shall discuss recent findings on the structure of
bio-molecular networks and discuss several graph models, including
{\it Scale-Free} graphs and {\it Duplication-Divergence} models,
that have been proposed to account for the properties observed in
real biological networks.  Section 4 is concerned with the
application of graph theoretical {\it measures of centrality or
importance} to biological networks. In particular, we shall
concentrate on the connection between the centrality of a gene or
protein within an interaction network and its likelihood to be
{\it essential} for the organism's survival. In Section 5, we
shall consider the {\it hierarchical structure} of biological
networks.  In particular, we shall discuss {\it motifs in
bio-molecular networks} and the identification of (typically
larger) functional modules.

In the second part of the article, we shall discuss two major
applications of Graph Theory to Biology.  Section 6 is concerned
with a number of issues and results related to the phenomenon of
synchronization in networks of inter-connected dynamical systems
and its relevance in various biological contexts.  Particular
attention will be given to suggested links between {\it patterns
of synchrony} and {\it neurological disorders}. In Section~7, we
shall discuss some recent work on the influence that the structure
of a social network can have on the behaviour of various disease
propagation models, and discuss the epidemiological significance
of these findings. Finally, in Section 8 we shall present our
concluding remarks and highlight some possible directions for
future research.

\section{Definitions and Mathematical Preliminaries}
The basic mathematical concept used to model networks is a {\it
graph}.  In this section, we shall introduce the principal
notations used throughout the paper, and recall some basic
definitions and facts from graph theory. While the material of
this section is mathematical in nature, we shall see in the
remainder of the paper that all of the concepts recalled here
arise in real biological networks. Furthermore, the notation and
nomenclature introduced in this section will enable us to discuss
the various biological networks encountered throughout the paper
in a uniform and consistent manner.

Throughout, $\RR$, $\RR^n$ and $\RR^{m \times n}$ denote the field
of real numbers, the vector space of $n$-tuples of real numbers
and the space of $m \times n$ matrices with entries in $\RR$
respectively.  $A^T$ denotes the transpose of a matrix $A$ in
$\RR^{m \times n}$ and $A \in \RR^{n \times n}$ is said to be
symmetric if $A = A^T$.

For finite sets $S, T$, $S \times T$ denotes the usual Cartesian
product of $S$ and $T$, while $|S|$ denotes the cardinality of
$S$.

\underline{\bf Directed and Undirected Graphs}

\begin{figure}[H]
             \begin{center}
              \psfrag{U}[tc][tc][1.675][0]{$u$}
              \psfrag{V}[tc][tc][1.675][0]{$v$}
              \resizebox{8cm}{!}{\includegraphics{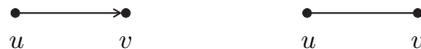}}
              \caption{An example of a directed graph (left) and an undirected graph (right), comprising two nodes and one edge.}
              \label{fig:directedundirected}
              \end{center}
\end{figure}

The concept of a {\it graph }is fundamental to the material to be
discussed in this paper.  The graphs or networks which we shall
encounter can be divided into two broad classes: {\it directed
graphs} and {\it undirected graphs}, as illustrated in
Figure~\ref{fig:directedundirected}.

Formally, a finite {\it directed graph}, $G$, consists of a set of
{\it vertices }or {\it nodes, }$\mathcal{V}(G)$,
\begin{eqnarray*}
\mathcal{V}(G) = \{v_1, \ldots, v_n \},
\end{eqnarray*}
together with an {\it edge }set, $\mathcal{E}(G) \subseteq
\mathcal{V}(G) \times \mathcal{V}(G)$.  Intuitively, each edge
$(u, v) \in \mathcal{E}(G)$ can be thought of as connecting the
starting node $u$ to the terminal node $v$.  For notational
convenience, we shall often write $uv$ for the edge $(u, v)$. We
shall say that the edge $uv$ {\it starts} at $u$ and {\it
terminates} at $v$.  For the most part, we shall be dealing with
graphs with finitely many vertices and for this reason, we shall
often omit the adjective finite where this is clear from context.

In Biology, transcriptional regulatory networks and metabolic
networks would usually be modelled as directed graphs.  For
instance, in a transcriptional regulatory network, nodes would
represent genes with edges denoting the interactions between them.
This would be a directed graph because, if gene A regulates gene
B, then there is a natural direction associated with the edge
between the corresponding nodes, starting at A and finishing at B.
Directed graphs also arise in the study of neuronal networks, in
which the nodes represent individual neurons and the edges
represent synaptic connections between neurons.

An {\it undirected graph}, $G$, also consists of a vertex set,
$\mathcal{V}(G)$, and an edge set $\mathcal{E}(G)$.  However,
there is no direction associated with the edges in this case.
Hence, the elements of $\mathcal{E}(G)$ are simply two-element
subsets of $\mathcal{V}(G)$, rather than ordered pairs as above.
As with directed graphs, we shall use the notation $uv$ (or $vu$
as direction is unimportant) to denote the edge $\{u, v \}$ in an
undirected graph. For two vertices, $u, v$ of an undirected graph,
$uv$ is an edge if and only if $vu$ is also an edge.  We are not
dealing with multi-graphs \cite{Die00}, so there can be at most
one edge between any pair of vertices in an undirected graph.  The
number of vertices $n$ in a directed or undirected graph is the
{\it size } or {\it order } of the graph.

In recent years, much attention has been focussed on the
protein-protein interaction networks of various simple organisms
\cite{Itoetal01, Raietal01}.  These networks describe the direct
physical interactions between the proteins in an organism's
proteome and there is no direction associated with the
interactions in such networks. Hence, PPI networks are typically
modelled as undirected graphs, in which nodes represent proteins
and edges represent interactions.

An edge, $uv$ in a directed or undirected graph $G$ is said to be
{\it an edge at} the vertices $u$ and $v$, and the two vertices
are said to be {\it adjacent }to each other. In this case, we also
say that $u$ and $v$ are {\it neighbours }. For an undirected
graph, $G$ and a vertex, $u \in \mathcal{V}(G)$, the set of all
neighbours of $u$ is denoted $\mathcal{N}(u)$ and given by
\begin{eqnarray*}
\mathcal{N}(u) = \{ v \in \mathcal{V}(G) : uv \in \mathcal{E}(G)
\}.
\end{eqnarray*}

\underline{\bf Node-degree and the Adjacency Matrix}

For an undirected graph $G$, we shall write ${\rm deg}(u)$ for the
degree of a node $u$ in $\mathcal{V}(G)$.  This is simply the
total number of edges at $u$.  For the graphs we shall consider,
this is equal to the number of neighbours of $u$,
\begin{eqnarray*} {\rm deg}(u) = | \mathcal{N}(u) |.
\end{eqnarray*}  In a directed graph $G$, the {\it in-degree} ,
${\rm deg}_\text{in}(u)$ ({\it out-degree }, ${\rm
deg}_\text{out}(u)$) of a vertex $u$ is given by the number of
edges that terminate (start) at $u$.

\begin{figure}[t]
  \centering
  \psfrag{U}[tc][tc][1.68][0]{$u$}
  \psfrag{A}[ll][ll][1.68][0]{$A=\begin{bmatrix}
                                0 & 1 & 1 & 1 \\
                                1 & 0 & 0 & 0 \\
                                1 & 0 & 0 & 0 \\
                                1 & 0 & 0 & 0
                           \end{bmatrix}$}
  \psfrag{B}[rr][rr][1.68][0]{$A=\begin{bmatrix}
                                0 & 1 & 1 & 0 \\
                                0 & 0 & 0 & 0 \\
                                0 & 0 & 0 & 0 \\
                                1 & 0 & 0 & 0
                           \end{bmatrix}$}
  \resizebox{12.8cm}{!}{\includegraphics{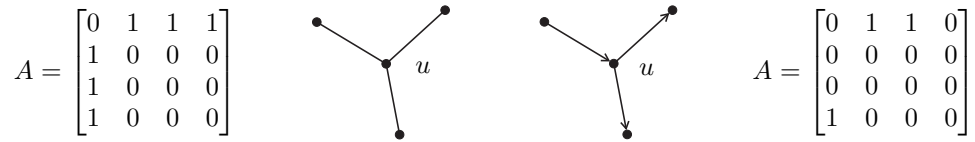}}
  \caption{The adjacency matrix of an undirected graph is
  symmetric; that of a directed graph generally is not. In this
example, we have that $\text{deg}(u)=3$ for the undirected graph
and $\text{deg}_\text{in}(u)=1$, $\text{deg}_\text{out}(u)=2$ for
the directed graph.}
  \label{fig:degree}
\end{figure}

Suppose that the vertices of a graph (directed or undirected) $G$
are ordered as $v_1, \ldots , v_n$.  Then the adjacency matrix,
$A$, of $G$ is given by
\begin{eqnarray}
\label{eq:adjmat} a_{ij} = \begin{cases} 1 & \mbox{ if } v_iv_j
\in \mathcal{E}(G) \\ 0 & \mbox{ if } v_iv_j \notin \mathcal{E}(G)
\end{cases}
\end{eqnarray}
Thus, the adjacency matrix of an undirected graph is symmetric
while this need not be the case for a directed graph.
Figure~\ref{fig:degree} illustrates this.

\underline{\bf Paths, Path Length and Diameter}

\begin{figure}[t]
  \centering
  \psfrag{A}[tr][tr][1.675][0]{$u=v_1$}
  \psfrag{B}[tc][tc][1.675][0]{$v_2$}
  \psfrag{C}[tc][tc][1.675][0]{$v_3$}
  \psfrag{D}[tc][tc][1.675][0]{$v_4$}
  \psfrag{E}[tl][tl][1.675][0]{$v_5=v$}
  \resizebox{8cm}{!}{\includegraphics{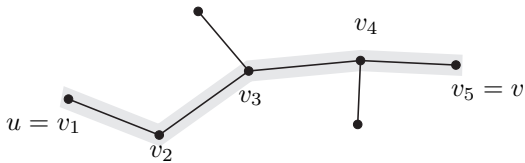}}
  \caption{A path of length $4$.}
  \label{fig:path}
\end{figure}

Let $u, v$ be two vertices in a graph $G$.  Then a sequence of
vertices
\begin{eqnarray*}
u = v_1, v_2, \ldots , v_k = v,
\end{eqnarray*}
such that for $i = 1, \ldots , k-1$:
\begin{itemize}
\item[(i)] $v_iv_{i+1} \in  \mathcal{E}(G)$; \item[(ii)] $v_i \neq
v_j$ for $i \neq j$
\end{itemize}
is said to be a path of length $k-1$ from $u$ to $v$.
Figure~\ref{fig:path} contains an example of a path of length $4$.

The {\it geodesic distance}, or simply distance, $\delta(u, v)$,
from $u$ to $v$ is the length of the shortest path from $u$ to $v$
in $G$.  If no such path exists, then we set $\delta(u, v) =
\infty$. If for every pair of vertices, $u, v \in \mathcal{V}(G)$,
there is some path from $u$ to $v$, then we say that $G$ is {\it
connected}.  The average path length and diameter of a graph $G$
are defined to be the average and maximum value of $\delta(u, v)$
taken over all pairs of distinct nodes, $u, v$ in $\mathcal{V}(G)$
which are connected by at least one path.

\underline{\bf Clustering Coefficient}

Suppose $u$ is a node of degree $k$ in an undirected graph $G$ and
that there are $e$ edges between the $k$ neighbours of $u$ in $G$.
Then the clustering coefficient of $u$ in $G$ is given by
\begin{eqnarray}
\label{eq:cluscoeff} C_u = \frac{2 e}{k(k-1)}.
\end{eqnarray}
Thus, $C_u$ measures the ratio of the number of edges between the
neighbours of $u$ to the total possible number of such edges,
which is $k(k-1)/2$.  The average clustering coefficient of a
graph $G$ is defined in the obvious manner.

\underline{\bf Statistical Notations}

Throughout the paper, we shall often be interested in average
values of various quantities where the average is taken over all
of the nodes in a given network of graph.  For some quantity, $f$,
associated with a vertex, $v$, the notation $\langle f \rangle$
denotes the average value of $f$ over all nodes in the graph.

\section{Identification and Modelling of Bio-molecular
Networks}\label{sec:models} As mentioned in Section
\ref{sec:intro}, this review paper naturally splits into two
parts. The first part consists of the current section and the
following two sections and is primarily focussed on the structural
properties of bio-molecular networks and on techniques that have
been developed for their analysis.

Due to recent advances in high-throughput technologies for
biological measurement, there is now more data available on
bio-molecular networks than ever before. This has made it possible
to study such networks on a scale which would have been impossible
two decades ago.  In fact, large-scale maps of protein interaction
networks \cite{Xenetal00, Mewetal02, Meretal02, Gioetal03,
Raietal01, Lietal04}, metabolic networks \cite{Jeoetal00,
Oveetal00} and transcriptional regulatory networks
\cite{Leeetal02, Tonetal04} have been constructed for a number of
simple organisms.  Motivated by these developments, there has been
a significant amount of work done on identifying and interpreting
the key structural properties of these networks in recent years.
In the current section, we shall give an overview of the main
aspects of this work.  In particular, we shall describe the
principal graph theoretical properties of bio-molecular networks
which have been observed in experimental data.  We shall also
discuss several mathematical models that have been proposed to
account for the observed topological properties of these networks.

\subsection{Structural Properties of Biological Networks}
In this subsection, we shall concentrate on the following three
aspects of network structure, which have received most attention
in the last few years:
\begin{itemize}
\item[(i)] Degree distributions; \item[(ii)] Characteristic path
lengths; \item[(iii)] Modular structure and local clustering
properties.
\end{itemize}
For each of these, we shall describe recently reported findings
for protein interaction, metabolic and transcriptional regulatory
networks in a variety of organisms.

\underline{\bf Degree Distributions}

Much of the recent research on the structure of bio-molecular and
other real networks has focussed on determining the form of their
degree distributions, $P(k), k = 0, 1, \ldots$, which measures the
proportion of nodes in the network having degree $k$.  Formally,
\begin{eqnarray*}
P(k) = \frac{n_k}{n},
\end{eqnarray*}
where $n_k$ is the number of nodes in the network of degree $k$
and $n$ is the size of the network.  It was reported in
\cite{FalFalFal99, BarAlb99} that the degree distributions of the
Internet and the WWW are described by a broad-tailed power law of
the form\footnote{In fact, the form $P(k) \sim (k + k_0)^{-
\gamma} e^{-k/k_c}$ with offset $k_0$ and an exponential cutoff
$k_c$ is more usually fitted to real network data.},
\begin{eqnarray}
\label{eq:PowLaw} P(k) \sim k^{- \gamma}, \quad\;\; \gamma > 1
\end{eqnarray}
Networks with degree distributions of this form are now commonly
referred to as {\it scale-free networks}.  This finding initially
surprised the authors of these papers as they had expected to find
that the degree distributions were Poisson or Gaussian.  In
particular, they has expected that the degrees of most nodes would
be close to the mean degree, $\langle k \rangle$, of the network,
and that $P(k)$ would decay exponentially as $| k- \langle k
\rangle |$ increased.  For such networks, the mean degree can be
thought of as typical for the overall network.  On the other hand,
the node-degrees in networks with broad-tailed distributions vary
substantially from their mean value, and $\langle k \rangle$
cannot be thought of as a typical value for the network in this
case.

Following on from the above findings on the WWW and the Internet,
several authors have investigated the form of the degree
distributions, $P(k)$, for various biological networks.  Recently,
several papers have been published that claim that interaction
networks in a variety of organisms are also scale-free.  For
instance, in \cite{Jeoetal00}, the degree distributions of the
central metabolic networks of 43 different organisms were
investigated using data from the WIT database \cite{Oveetal00}.
The results of this paper indicate that, for all 43 networks
studied, the distributions of in-degree, $P_{in}(k)$, and
out-degree, $P_{out}(k)$, have tails of the form
(\ref{eq:PowLaw}), with $2 < \gamma < 3$.

Similar studies on the degree distributions of protein interaction
networks in various organisms have also been carried out.  In
\cite{YooOltBar04}, the protein interaction network of {\it S.
cerevisiae} was analysed using data from four different sources.
As is often the case with data of this nature, there was little
overlap between the interactions identified in the different sets
of data.  However, in all four cases, the degree distribution
appeared to be broad-tailed and to be best described by some form
of modified power law.  Similar findings have also been reported
for the protein interaction networks of {\it E. coli, D.
melanogaster, C. elegans} and {\it H. pylori} in the recent paper
\cite{GohKahKim05}. Note however that for transcriptional
regulatory networks, while the outgoing degree distribution again
appears to follow a power law, the incoming degree distribution is
better approximated by an exponential rule of the form $P_{in}(k)
\sim e^{-\beta k}$ \cite{BarOlt04, Gueetal02, FeaBro02}.

At this point, it is important to record some remarks on the
observations of scale-free topologies in biological interaction
networks.  First of all, the broad-tailed degree distributions
observed in these networks is not consistent with the traditional
random graph models which have been used to describe complex
networks \cite{Bol01, AlbBar02}. In these models, node-degrees are
closely clustered around the mean degree, $\langle k \rangle$, and
the probability of a node having degree $k$ decreases
exponentially with $| k - \langle k \rangle |$. However, in
scale-free networks, while most nodes have relatively low degree,
there are significant numbers of nodes with unusually high degree
- far higher than the mean degree of the network. Such nodes are
now usually referred to as {\it hubs. } It has been noted
\cite{AlbJeoBar00} that the scale-free structure has implications
for the robustness and vulnerability of networks to failure and
attack.  Specifically, while removing most of the nodes in a
scale-free network will have little effect on the network's
connectivity, the targeted removal of hub nodes can disconnect the
network relatively easily. This has led to the suggestion that
genes or proteins which are involved in a large number of
interactions, corresponding to hub nodes, may be more important
for an organism's survival than those of low degree. The
connection between network topology and the biological importance
of genes and proteins has been extensively studied recently and we
shall describe this strand of research in detail in Section
\ref{sec:Centrality}.

A second important point is that all of the analysis described
above has been carried out on {\it sampled subnetworks }rather
than on a complete network.  For instance, the protein interaction
networks which have been studied usually contain only a fraction
of the complete set of proteins of an organism.  Moreover, the
interactions included in these networks are far from complete.
Thus, the conclusion being drawn are based on a subnetwork
containing only a sample of the nodes and edges of the complete
network. While some studies have indicated that the statistical
properties of interaction networks may be robust with respect to
variations from one data set to another, the impact of sampling
and inaccurate/incomplete information on the identified degree
distributions is an important issue which is not yet fully
understood.  For instance, in \cite{Thoetal03} it was shown using
a model of protein interaction networks that an approximate power
law distribution can be observed in a sampled sub-network while
the degree distribution of the overall network is quite different.
Further evidence of the need for caution in drawing conclusions
about the overall structure of biological networks based on
samples has been provided in \cite{ClaMoo05, StuWiuMay05}, where
results on the sampling properties of various types of network
models were presented.  For instance, in \cite{ClaMoo05}, a
sampling regime based on the construction of {\it spanning trees}
\cite{Die00} was studied.  Here, starting from a source vertex
$v_0$, a tree $T$ is constructed by first adding the neighbours of
$v_0$ to $T$ and then selecting one of these, and repeating the
process.  In this paper, approximate arguments were presented to
show that such a sampling regime can lead to a subnetwork with
degree distribution of the form $P(k) \sim 1/k$ even when the
complete network has a Poisson degree distribution.

\underline{\bf Diameter and Characteristic Path Length}

Several recent studies have revealed that the average path lengths
and diameters of bio-molecular networks are ``small'' in
comparison to network size.  Specifically, if the size of a
network is $n$, the average path length and diameter are of the
same order of magnitude as ${\rm log }(n)$ or even smaller.  This
property has been previously noted for a variety of other
technological and social networks \cite{AlbBar02}, and is often
referred to as the {\it small world property} \cite{WatStr98}.
This phenomenon has now been observed in metabolic, genetic and
protein interaction networks.  For instance, in \cite{WagFel01,
Jeoetal00}, the average path lengths of metabolic networks were
studied.  The networks analysed in these papers had average path
lengths between 3 and 5 while the network sizes varied from
200-500. Similar findings have been reported for genetic networks
in \cite{Tonetal04}, where a network of approximately 1000 genes
and 4000 interactions was found to have a characteristic path
length of 3.3, and for protein interaction networks in
\cite{Wag01, Yuetal04, YooOltBar04}.

In a sense, the average path length in a network is an indicator
of how readily ``information'' can be transmitted through it.
Thus, the small world property observed in biological networks
suggests that such networks are efficient in the transfer of
biological information: only a small number of intermediate
reactions are necessary for any one protein/gene/metabolite to
influence the characteristics or behaviour of another.

\underline{\bf Clustering and Modularity}

The final aspect of network structure which we shall discuss here
is concerned with how densely clustered the edges in a network
are.  In a highly clustered network, the neighbours of a given
node are very likely to be themselves linked by an edge.
Typically, the first step in studying the clustering and modular
properties of a network is to calculate its average clustering
coefficient, $C$, and the related function, $C(k)$, which gives
the average clustering coefficient of nodes of degree $k$ in the
network.  As we shall see below, the form of this function can
give insights into the global network structure.

In \cite{Ravetal02}, the average clustering coefficient was
calculated for the metabolic networks of 43 organisms and, in each
case, compared to the clustering coefficient of a random network
with the same underlying degree distribution.  In fact, the
comparison was with the Barabasi-Albert (BA) model of scale-free
networks which we shall discuss in the next subsection.  In each
case, the clustering coefficient of the metabolic network was at
least an order of magnitude higher than that of the corresponding
BA network.  Moreover, the function $C(k)$ appeared to take the
form $C(k) \sim k^{-1}$.  Thus, as the degree of a node increases,
its clustering coefficient decreases.  This suggests that the
neighborhoods of low-degree nodes are densely clustered while
those of hub nodes are quite sparsely connected.  In order to
account for this, the authors of \cite{Ravetal02} suggested a
hierarchical modular structure for metabolic networks in which:
\begin{itemize}
\item[(i)]  Individual modules are comprised of densely clustered
nodes of relatively low degree; \item[(ii)] Different modules are
linked by hub nodes of high degree.
\end{itemize}
Similar results for the clustering coefficient and the form of the
function $C(k)$ have been reported in \cite{GohKahKim05} for the
protein interaction networks of {\it S. cerevisiae, H. pylori, E.
coli} and {\it C. elegans}, indicating that these undirected
networks may also have a modular structure, in which hub nodes act
as links or bridges between different modules within the networks.
Further evidence for the intermediary role of hub nodes was
provided in \cite{MasSne02} where correlations between the degrees
of neighbouring nodes in the protein interaction network and the
transcriptional regulatory network of {\it S. cerevisiae} were
investigated.  The authors of this paper found clear evidence of
such correlation; in fact, for both networks, nodes of high degree
are significantly more likely to connect to nodes of low degree
than to other ``hubs''.  This property of a network is referred to
as {\it disassortativity. }For more discussion on this topic, see
\cite{Coletal05}. Finally, we note that a high degree of local
clustering has also been observed in the transcriptional
regulatory network of {\it S. cerevisiae} in \cite{Tonetal04}.

\subsection{Mathematical Models for Interaction Networks}
Given the empirically observed properties of interaction networks
discussed above, it is natural to ask whether these can be
explained by means of mathematical models based on plausible
biological assumptions.  Furthermore:
\begin{itemize}
\item[(i)] Reliable models for the evolution of interaction
networks may deepen our understanding of the biological processes
behind their evolution. \item[(ii)] Such models could be used to
assess the reliability of experimental results on network
structure and to assist in experimental design.  For instance, the
strategy for optimally identifying protein-protein interaction
(PPI) network structure described in \cite{LapHol04} relies on the
statistical abundance of nodes of high degree in scale-free
networks which we shall discuss in more detail below. Furthermore,
this strategy was suggested as a means of determining the PPI
network in humans. Note also the work described in \cite{GolRot03}
on assessing the reliability of network data and predicting the
existence of links in a PPI network which have not yet been
determined.  The methods in this paper were based on properties of
the {\it small-world }network model of Watts and Strogatz
introduced in \cite{WatStr98} to described social and neurological
networks.
\end{itemize}
To date, several different mathematical models of complex networks
have been proposed in the literature.  A number of these were not
developed with specifically biological networks in mind, but
rather to account for some of the topological features observed in
real networks in Biology and elsewhere.  On the other hand, in the
recent past several models for protein interaction and genetic
networks have been proposed based on biological assumptions.  In
this subsection, we shall describe the main models that have been
used to model biological networks, and some theoretical results on
the structure of these models.

\underline{\bf Classical Models and Scale-free Graphs}

In the 1950's,  Paul Erd\"{o}s and Alfred Renyi introduced their
now classical notion of a {\it random graph} to model non-regular
complex networks.  The basic idea of the Erd\"{o}s-Renyi (ER)
random graph model is the following. Let a set of $n$ nodes,
$\{v_1, \ldots , v_n \}$, and a real number $p$ with $0 \leq p
\leq 1$ be given.  Then for each pair of nodes, $v_i, v_j$, an
edge is placed between $v_i$ and $v_j$ with probability $p$.
Effectively, this defines a probability space where the individual
elements are particular graphs on $\{v_1, \ldots, v_n \}$ and the
probability of a given graph with $m$ edges occurring is $p^m
(1-p)^{n-m}$.  For background on the mathematical theory of ER
graphs, consult \cite{Bol01, Die00}.

The theory of random graphs has been a highly active field of
mathematics for fifty years and many deep theorems about the
properties of ER graphs have been established.  For example, it
has been proven for these networks that the characteristic path
length is proportional to the logarithm of the network size, and
that the average clustering coefficient is inversely proportional
to network size.  Perhaps the most relevant fact about the ER
model in our context is the relatively straightforward result that
the degree distribution is binomial. Thus, the degree distribution
of a large ER network can be approximated by a Poisson
distribution.  The tails of such distributions are typically
narrow, meaning that, for ER graphs, the node degrees tend to be
tightly clustered around the mean degree $\langle k \rangle$.

This last fact contrasts with the findings reported in the
previous subsection that the degree distributions of many
biological networks appear to follow a broad-tailed power law. The
same observation has also been made for several man-made networks
including the WWW and the Internet. This behaviour is inconsistent
with the classical ER model of random graphs and led Barabasi,
Albert and co-workers to devise a new model for the dynamics of
network evolution.  This model is based on the two fundamental
mechanisms of {\it growth }and {\it preferential attachment}, and
has been the subject of intensive research in the last few years.
It is usually referred to as the Barabasi-Albert (BA) model.

The core idea of Barabasi and Albert was to consider a network as
an evolving entity and to model the dynamics of network growth.
The simple BA model is now well known and is usually described in
the following manner \cite{AlbBar02}. Given a positive integer,
$m$ and an initial network, $G_0$, the network evolves according
to the following rules (note that this is a discrete-time
process):
\begin{itemize}
\item[(i)] {\it Growth:} At each time $j$, a new node of degree
$m$ is added to the network; \item[(ii)] {\it Preferential
Attachment:} For each node $u$ in the existing network, the
probability that the new node connects to it is proportional to
the degree of $u$.  Formally, writing $G_j$ for the network at
time $j$ and $P(u, j)$ for the probability that the new node added
at time $k$ is linked to $u$ in $G_{j-1}$:
\begin{eqnarray}
\label{eq:PrefAtt} P(u, j) = \frac{{\rm deg}(u)}{\sum_{v \in
\mathcal{V}(G_{j-1})} {\rm deg}(v)}.
\end{eqnarray}
\end{itemize}

Using computer simulations and approximate arguments based on
``mean field theory'' it has been argued that the above scheme
generates a network whose degree distribution asymptotically
approaches the power law $P(k) \sim k^{-\gamma}$ with $\gamma = 3$
\cite{AlbBar02}. A number of variations on the basic BA model have
also been proposed that have power law degree distributions with
values of the degree exponent other than three.  See for instance,
the models for evolving networks described in \cite{DorMenSam00,
KraRedLey00} which give rise to power law degree distributions
with exponents in the range $2 < \gamma < + \infty$.

{\it Some Issues in the Use of Scale-free Models}

While the degree distributions of BA and related scale-free models
appear to fit the experimental data on bio-molecular networks more
accurately than classical ER networks, there are several issues
related to their use that should be noted. In \cite{Boletal01}, it
was pointed out that the commonly used definition of BA graphs is
ambiguous.  For instance, the question of how to initiate the
process of network evolution is not explicitly dealt with in the
original papers; how do we connect the new node to the existing
nodes with probability proportional to their degrees if all such
nodes have degree zero to begin with?  This issue can be
circumvented by beginning with a network which has no isolated
nodes.  However, this immediately raises the difficult question of
how the choice of initial network influences the properties of the
growing network.  These issues have been discussed in detail in
\cite{Boletal01, BolRio02} where more mathematically rigorous
formulations of the preferential attachment mechanism for network
growth have been presented. A number of formal results concerning
degree distributions, network diameter, robustness to node removal
and other network properties have also been presented in
\cite{BolRio03, BolRio04}.

There has been a remarkable level of interest in the scale-free
family of random graphs in recent years and numerous papers
claiming that biological interaction networks follow a power law
and belong to this class have been published. However, it is
important to note that a number of reservations about the use of
the BA and related models in Biology have been raised recently
\cite{Fox05}.
\begin{itemize}
\item[(i)] Firstly, the BA model is not based on specific
biological considerations.  Rather, it is a mathematical model for
the dynamical growth of networks that replicates the degree
distributions, and some other properties, observed in studies of
the WWW and other networks.  In particular, it should be kept in
mind that the degree distribution is just one property of a
network and that networks with the same degree distribution can
differ substantially in other important structural aspects
\cite{VolVolBla02}. \item[(ii)] Many of the results on BA and
related networks have only been empirically established through
simulation, and a fully rigorous understanding of their properties
is still lacking. A number of authors have started to address this
issue in the recent past but this work is still in an early stage.
Also, as noted above, the definition of BA graphs frequently given
in the literature is ambiguous \cite{Boletal01}. \item[(iii)] Most
significantly, from a practical point of view, the observations of
scale-free and power law behaviour in biological networks are
based on partial and inaccurate data.  The techniques used to
identify protein interactions and transcriptional regulation are
prone to high levels of false positive and false negative errors
\cite{Meretal02}. Moreover, the networks being studied typically
only contain a fraction of the genes or proteins in an organism.
Thus, we are in effect drawing conclusions about the topology of
an entire network based on a {\it sample }of its nodes, and a
noisy sample at that. In order to do this reliably, a thorough
understanding of the effect of sampling on network statistics,
such as distributions of node degrees and clustering coefficients,
is required.  Some authors have recently started to address this
issue and the following two results, presented in \cite{ClaMoo05,
StuWiuMay05}, are extremely relevant in the present context:
\begin{itemize}
\item[(a)] Subnetworks sampled from a scale free network are not
in general scale free; \item[(b)] It is possible for a sampled
subnetwork of a network with Poisson degree distribution (which is
certainly not scale-free) to appear to be scale-free.
\end{itemize}
Further results of a similar nature have recently been reported in
\cite{Hanetal05}.  Here, the sampling process in the large-scale
yeast-2-hybrid (Y2H) experiments which have generated many of the
existing protein-interaction maps for yeast was simulated on four
different types of network models. The degree distributions of the
models considered were normal, exponential, scale-free and
truncated normal respectively.  Based on the findings reported in
this paper, the authors argued that, given the coverage of the
yeast interactome currently available, none of the four models
considered could be definitively ruled out as a model of the
complete yeast interaction network! These facts cast doubt on the
hypothesis that the complete PPI networks of living organisms are
in fact scale free. At the very least, it demonstrates the need to
be careful about the effects of sampling and data noise when we
attempt to draw conclusions about the structure of biological and
other real world networks.
\end{itemize}
Before moving on to discuss a number of more biologically
motivated models for interaction networks, we note the recent
paper \cite{PrzCorJur04} in which {\it geometric random graphs }
\cite{Pen03} were suggested as an alternative model for protein
interaction networks.  This suggestion was based on comparing the
frequency of small subgraphs in real networks to their frequency
in various network models, including geometric graphs.  However,
as with BA models, there is no clear biological motivation for
choosing geometric graphs to model protein interaction networks
and, furthermore, the comparisons presented in \cite{PrzCorJur04}
are based on a very small number of sample random networks.  On
the other hand, the authors of this paper make the important point
that the accuracy of network models is crucial if we are to use
these to assess the reliability of experimental data or in the
design of experiments for determining network structure.

\underline{\bf Duplication and Divergence Models}

Many of the recent models for network evolution are founded on
some variation of the basic mechanisms of growth and preferential
attachment.   However, there are other, more biologically
motivated models which have been developed specifically for
protein interaction and genetic regulatory networks.  As with the
models discussed above, these are usually based on two fundamental
processes: {\it duplication }and {\it divergence}. The hypothesis
underpinning these so-called Duplication-Divergence (DD) models is
that gene and protein networks evolve through the occasional
copying of individual genes/proteins, followed by subsequent
mutations. Over a long period of time, these processes combine to
produce networks consisting of genes and proteins, some of which,
while distinct, will have closely related properties due to common
ancestry.

To illustrate the main idea behind DD models, we shall give a
brief description of the model for protein interaction networks
suggested in \cite{Vazetal03}.  Given some initial network $G_0$,
the network is updated at each time $t$ according to the rules:
\begin{itemize}
\item[(i)] {\it Duplication:} A node $v$ is chosen from the
network $G_{t-1}$ at random and a new node $v'$ - a duplicate of
$v$ - is added to the network and connected to all of the
neighbours of $v$; \item[(ii)] {\it Divergence:} For each
neighbour, $w$, of $v'$, the edge $v'w$ is removed with
probability $q$.
\end{itemize}
As pointed out in \cite{Vazetal03}, the above scheme effectively
introduces a preferential attachment mechanism into the network
and generates a power law degree distribution.  A number of basic
properties of the model and its suitability to model the PPI
network of S. cerevisae are discussed in this paper also.  The
same basic model has also been studied more analytically in
\cite{Chuetal03}.  In this paper, it was shown that if $q < 1/2$,
then the degree distribution of the DD network is given by a power
law whose exponent $\gamma$ satisfies $\gamma < 2$. The authors of
this paper also considered some closely related models for the
growth of gene networks in the earlier paper \cite{BhaGalDew02}.
Here it was pointed out that duplication alone will not give rise
to a power law degree distribution.

The model described in \cite{Vazetal03} allowed for
self-interacting proteins, where the copy $v'$ can also form a
link to the original $v$ with some non-zero probability. However,
there are several assumptions associated with the basic scheme
described above whose biological validity is questionable.
\begin{itemize}
\item[(i)] The new node, $v'$, can only form links to neighbours
of the original node $v$ - this restricts the types of mutations
allowed for duplicate genes; \item[(ii)] A node can only undergo
mutation or divergence at the instant when it is added to the
network - this ignores the possibility of genes continuing to
mutate long after the duplication event; \item[(iii)] Nodes and
edges can only be added to the network and not removed - this
clearly places a significant restriction on the types of mutation
and evolution possible.
\end{itemize}
Several extensions of the basic DD model have been proposed to
relax some of the assumptions outlined above.  For instance, point
(i) above has been addressed in \cite{Soletal02}, while a model
that allows for edge additions and removals at a much faster rate
than gene duplications has been described and analysed in
\cite{BerLasWag04}.  Yet another growth model (based on a
preferential attachment mechanism) which allows edges to be added
and deleted between nodes in the existing network, and for new
nodes to be added to the network has been presented in
\cite{Wag02}. Finally, the issue in point (iii) has been addressed
in the recent paper \cite{ChuLu04} by a growth-deletion model that
allows for the addition and removal of both edges and nodes.

While DD models of network growth are based on more plausible
biological assumptions than the BA-type models, several of the
caveats expressed above for BA networks still apply.  In
particular, the question of how reliably we can infer a network's
structure from studying a sample of its nodes is critical, as is
the impact of noisy data on identifying network structure.
However, these points should not be seen as a criticism of the
models themselves.  Our aim is rather to highlight important
issues that need to be taken into account in assessing how
accurately such models reflect the biological reality.  Of course,
more reliable data is required for this to be possible. Finally,
we should note that the theory of DD networks is still in a very
early stage and many of their key mathematical properties are only
partially understood. 

\subsection{Summarizing Comments}
\begin{itemize}
\item[(i)] Significant progress has been made recently on
constructing maps of bio-molecular networks in simple organisms.
Using the available data, the structural properties of
protein-protein interaction, transcriptional regulatory and
metabolic networks have been studied and preliminary results have
been reported.  The networks studied appear to have scale-free
degree distributions, short characteristic path lengths and high
clustering coefficients.  The observed properties are not in
agreement with those of traditional random graph models for
complex networks.  \item[(ii)] Several new mathematical models for
the growth of random networks have been proposed in the recent
past. These include a number of variations on the basic BA
scale-free model, and the more biologically inspired {\it
Duplication-Divergence } models for gene and protein networks. The
mathematical theory of these models is only beginning to be
developed and offers many exciting and challenging opportunities
for future biologically motivated research.  \item[(iii)] Both in
the identification of network properties, and in the development
of mathematical models, the issues of inaccurate data and sampling
are of paramount importance.  Recent results on the sampling
properties of networks with power law and Poisson degree
distributions highlight the need for caution when drawing
conclusions on global network properties from an analysis of a
sampled subnetwork.
\end{itemize}

\section{Measures of Centrality and Importance in Biological
Networks}\label{sec:Centrality}

The problem of identifying the most important nodes in a large
complex network is of fundamental importance in a number of
application areas, including Communications, Sociology and
Management.  To date, several measures have been devised for
ranking the nodes in a complex network and quantifying their
relative importance.   Many of these originated in the Sociology
and Operations Research literature, where they are commonly known
as {\it centrality measures } \cite{WasFau94}. More recently,
driven by the phenomenal growth of the World Wide Web, schemes
such as the PageRank algorithm on which GOOGLE is based, have been
developed for identifying the most relevant web-pages to a
specific user query.

As described in the previous section, there is now a large body of
data available on bio-molecular networks, and there has been
considerable interest in studying the structure of these networks
and relating it to biological properties in the recent past. In
particular, several researchers have applied centrality measures
to identify structurally important genes or proteins in
interaction networks and investigated the biological significance
of the genes or proteins identified in this way.  Particular
attention has been given to the relationship between centrality
and essentiality, where a gene or protein is said to be essential
for an organism if the organism cannot survive without it.  The
use of centrality measures to predict essentiality based on
network topology has potentially significant applications to drug
target identification \cite{VogLanLev00, JeoOltBar03}.

In this section, we shall describe several measures of network
importance or centrality that have been applied to protein
interaction and transcriptional regulatory networks in the recent
past.  We shall place particular emphasis on the efforts to assess
the biological significance of the most central genes or proteins
within these networks.

\subsection{Classical Centrality Measures}
In this subsection, we shall discuss four classical concepts of
centrality which have recently been applied to biological
interaction networks:
\begin{itemize}
\item[(i)] Degree centrality; \item[(ii)] Closeness centrality;
\item[(iii)] Betweenness centrality; \item[(iv)] Eigenvector
centrality.
\end{itemize}
\underline{\bf Degree Centrality}

Degree centrality is the most basic of the centrality measures to
be discussed here. The idea behind using degree centrality as a
measure of importance in network is the following:
\begin{center}
\begin{quote}
{\it An important node is involved in a large number of
interactions.}
\end{quote}
\end{center}
Formally, for an undirected graph $G$, the degree centrality of a
node $u \in \mathcal{V}(G)$ is given by
\begin{eqnarray}
\label{eq:DegCen} C_d(u) = {\rm deg}(u).
\end{eqnarray}
For directed networks, there are two notions of degree centrality:
one based on in-degree and the other on out-degree. These are
defined in the obvious manner.  Degree centrality and the other
measures discussed here are often normalised to lie in the
interval $[0, 1]$.

As discussed in the previous section, a number of recent studies
have indicated that bio-molecular networks have broad-tailed
degree distributions, meaning that while most nodes in such
networks have a relatively low degree, there are significant
numbers of so-called hub nodes.  The removal of these hub nodes
has a far greater impact on the topology and connectedness of the
network than the removal of nodes of low degree
\cite{AlbJeoBar00}. This naturally leads to the hypothesis that
hub nodes in protein interaction networks and genetic regulatory
networks may represent essential genes and proteins.  In
\cite{Jeoetal01}, the connection between degree centrality and
essentiality was investigated for the protein-protein interaction
network in {\it S. cerevisiae}.  The analysis was carried out on a
network consisting of 1870 nodes connected by 2240 edges, which
was constructed by combining the results of earlier research
presented in \cite{Uetzetal00, Xenetal00}.  In this network,
$21\%$ of those proteins that are involved in fewer than 5
interactions, $C_d(u) \leq 5$, were essential while, in contrast,
$62\%$ of proteins involved in more than 15 interactions, $C_d(v)
\geq 15$, were essential.

More recently, similar findings were reported in \cite{Yuetal04}.
Again, the authors considered a network of protein interactions in
yeast, this time consisting of 23294 interactions between 4743
proteins. The average degree of an essential protein in this
network was 18.7, while the average degree of a non-essential
protein was only 7.4.  Moreover, defining a hub to be a node in
the first quartile of nodes ranked according to degree, the
authors of \cite{Yuetal04} found that over $40\%$ of hubs were
essential while only $20\%$ of all nodes in the network are
essential.

The above observations have led some authors to propose that, in
protein interaction networks, node degree and essentiality may be
related \cite{Yuetal04, Jeoetal01}.  However, the precise nature
of this relationship is far from straightforward.  For instance,
using a network constructed from data published in
\cite{Itoetal01, Uetzetal00}, the author of \cite{Wuc02} has
claimed that there is little difference between the distributions
of node degrees for essential and non-essential proteins in the
interaction network of yeast. However, in this network, the
degrees of essential proteins are still typically higher than
those of non-essential proteins.

In \cite{HahConWag04} the connection between the degree of a
protein and the rate at which it evolves was investigated.  The
authors reasoned that if highly connected proteins are more
important to an organism's survival, they should be subject to
more stringent evolutionary constraints and should evolve at a
slower rate than non-essential proteins.  However, the authors of
\cite{HahConWag04} found no evidence of a significant correlation
between the number of interactions in which a protein is involved
and its evolutionary rate.  Once again, this indicates that while
node degree gives some indication of a protein's likelihood to be
essential, the precise relationship between essentiality and node
degree is not a simple one.

\underline{\bf Closeness Centrality Measures}

We shall now discuss {\it closeness centrality measures } which
are defined in terms of the geodesic distance, $\delta(u, v)$
between nodes in a graph or network.  The basic idea behind this
category of measures is the following:
\begin{center}
\begin{quote}
{\it An important node is typically ``close'' to, and can
communicate quickly with, the other nodes in the network.}
\end{quote}
\end{center}

In the recent paper \cite{WucSta03}, three closeness measures,
which arise in the context of resource allocation problems, were
applied to metabolic and protein interaction networks. The
specific measures considered in this paper were {\it excentricity,
status, and centroid value}.

The excentricity, $C_e(u)$, of a node $u$ in a graph $G$ is given
by
\begin{eqnarray}
\label{eq:e(v)} C_e(u) = \max_{v \in \mathcal{V}(G)} \delta(u, v),
\end{eqnarray}
and the {\it centre }of $G$ is then the set
\begin{eqnarray}
\label{eq:C(G)} \mathcal{C}(G) = \{ v \in \mathcal{V}(G) : C_e(v)
= \min_{w \in \mathcal{V}(G)} C_e(w) \}.
\end{eqnarray}
Thus, the nodes in $\mathcal{C}(G)$ are those that minimise the
maximum distance to any other node of $G$.

The {\it status}, $C_s(u)$, of a node $v$ is given by
\begin{eqnarray}
\label{eq:s(v)} C_s(u) = \sum_{v \in \mathcal{V}(G)} \delta(u, v),
\end{eqnarray}
and the {\it median }of $G$ is then the set
\begin{eqnarray}
\label{eq:M(G)} \mathcal{M}(G) = \{ v \in \mathcal{V}(G) : C_s(v)
= \min_{w \in \mathcal{V}(G)} C_s(w) \}.
\end{eqnarray}
The nodes in $\mathcal{M}(G)$ minimise the {\it average }distance
to other nodes in the network.

The final measure considered in \cite{WucSta03} is the {\it
centroid }value which is closely related to the status defined
above.  In fact, these two measures give rise to identical
rankings of the nodes in a graph and, for this reason, we shall
not formally define centroid value here.

A number of points about the results presented in \cite{WucSta03}
are worth noting.  First of all, on both ER graphs and the BA
model of scale-free graphs, all three measures were found to be
strongly correlated with node-degree.  The measures were then
applied to the central metabolic network of {\it E. coli} and the
centre, $\mathcal{C}(G)$, and the median, $\mathcal{M}(G)$, of
this network were calculated.  The authors reasoned that central
nodes represent ``cross-roads'' or ``bottlenecks'' in a network
and should correspond to key elements of the organism's
metabolism. In support of this assertion, the centre,
$\mathcal{C}(G)$, contained several of the most important known
substrates, including ATP, ADP, AMP and NADP.  On the other hand,
in the protein interaction network of {\it S. cerevisiae}, no
discernible difference between the excentricity distribution of
essential and non-essential proteins was observed.   In the same
paper, centrality measures were also applied to networks of
protein domains where two domains are connected by an edge if they
co-occur in the same protein. The nodes with the highest
centrality scores in these networks corresponded to domains
involved in signal transduction and cell-cell contacts.

\underline{\bf Betweenness Centrality Measures}

In \cite{Fre78}, the concept of {\it betweenness centrality }was
introduced as a means of quantifying an individual's influence
within a social network.  The idea behind this centrality measure
is the following:
\begin{center}
\begin{quote}
{\it An important node will lie on a high proportion of paths
between other nodes in the network.}
\end{quote}
\end{center}
Formally, for distinct nodes, $u, v, w \in \mathcal{V}(G)$, let
$\sigma_{uv}$ be the total number of geodesic paths between $u$
and $v$ and $\sigma_{uv}(w)$ be the number of geodesic paths from
$u$ to $v$ that pass through $w$.  Also, for $w \in
\mathcal{V}(G)$, let $V(u)$ denote the set of all ordered pairs,
$(u, v)$ in $\mathcal{V}(G) \times \mathcal{V}(G)$ such that $u,
v, w$ are all distinct. Then, the betweenness centrality of $w$,
$C_b(w)$, is given by
\begin{eqnarray}
\label{eq:BetCen} C_b(w) = \sum_{(u, v) \in V(w)}
\frac{\sigma_{uv}(w)}{\sigma_{uv}}.
\end{eqnarray}
Recently, in \cite{Joyetal05} the measure $C_b$ was applied to the
yeast protein interaction network and the mean value of $C_b$ for
the essential proteins in the network was approximately $80\%$
higher than for non-essential proteins.  In fact, the results in
this paper indicate that the performance of $C_b$ as an indicator
of essentiality is comparable to that of node degree.  In this
paper, it was also noted that there were significant numbers of
proteins with high betweenness centrality scores but low node
degree.  The authors pointed out that this was not consistent with
the scale-free BA model or with the more biologically motivated DD
models proposed in \cite{Soletal02, Vazetal03}.  Furthermore,
there was considerable variation in the value of $C_b(u)$ for
proteins $u$ with the same degree.  This naturally raises the
following question: if two proteins, $u$, $v$ have the same degree
$k$ but $C_b(u) > C_b(v)$, is $u$ more likely to be essential than
$v$? However, no clear evidence to support this hypothesis was
found in the data considered in \cite{Joyetal05}.

In the present context, it is worth noting the work in
\cite{New05} where a definition of betweennness centrality based
on random paths between nodes, rather than on geodesic paths was
considered. This centrality measure was motivated by the fact
that, in real networks, information does not always flow along the
shortest available path between two points.  To the best of the
authors' knowledge, this new concept of betweenness centrality has
yet to be applied to bio-molecular networks in a systematic way.

\underline{\bf Eigenvector Centrality Measures}

As with many of the measures considered in this section,
eigenvector centrality measures appear to have first arisen in the
analysis of social networks, and several variations on the basic
concept described here have been proposed \cite{Bon72, Bon87,
WasFau94, BonLlo01}.  This family of measures is a little more
complicated than those considered previously and eigenvector
centrality measures are usually defined as the limits of some
iterative process. The core idea behind these measures is the
following.
\begin{center}
\begin{quote}
{\it An important node is connected to important neighbours.}
\end{quote}
\end{center}
In much of the original work presented in the sociology
literature, the eigenvector centrality scores of a network's nodes
were determined from the entries of the principal eigenvector of
the network's adjacency matrix. Formally, if $A$ is the adjacency
matrix of a network $G$ with $\mathcal{V}(G) = \{ v_1, \ldots ,
v_n\}$, and \[ \rho(A) = \max_{\lambda \in \sigma(A)} |\lambda|\]
is the spectral radius of $A$, then the eigenvector centrality
score, $C_{ev}(v_i)$ of the node $v_i$ is given by the $i^{th}$
co-ordinate, $x_i$, of a suitably normalised eigenvector $x$
satisfying
\begin{eqnarray*}
Ax = \rho(A) x.
\end{eqnarray*}
In the recent paper \cite{Est05}, the connection between various
centrality measures, including eigenvector centrality, and
essentiality within the protein interaction network of yeast was
investigated.  In this paper, the performance of eigenvector
centrality was comparable to that of degree centrality and it
appeared to perform better than either betweenness centrality or
closeness centrality measures.  A number of other centrality
measures which we shall mention later in this section were also
studied.  Before concluding our discussion of the classical
centrality measures and their possible application to the
identification of essential genes or proteins, it is worth noting
the following points about eigenvector centrality.
\begin{itemize}
\item[(i)] In order for the definition above to uniquely specify a
ranking of the nodes in a network it is necessary that the
eigenvalue $\rho(A)$ has geometric multiplicity one.  For general
networks, this need not be the case.  However, if the network is
connected then it follows from the Perron-Frobenius Theorem for
irreducible non-negative matrices \cite{BerPle94, HorJohn85} that
this will be the case. \item[(ii)] Similar ideas to those used in
the definition of eigenvector centrality have recently been
applied to develop the Page-Rank algorithm on which the GOOGLE
search engine relies \cite{BriPag98, LanMey05}. The HITS algorithm
for the ranking of web pages, proposed by Kleinberg \cite{Kle98},
also relies on similar reasoning.
\end{itemize}

\underline{\bf Other Centrality Measures}

Finally for this subsection, we briefly note several less standard
centrality measures which have been developed in the last decade
or so, with potential applications in the analysis of biological
networks.  For instance, in \cite{Est05} the notion of {\it
subgraph centrality }was introduced and the relationship between
the subgraph centrality of a protein in the yeast interaction
network and its likelihood to be essential was investigated.
Loosely speaking, the subgraph centrality of a node measures the
number of subgraphs of the overall network in which the node
participates, with more weight being given to small subgraphs.
Formally, if $A$ is the adjacency matrix of a network with vertex
set, $\mathcal{V}(G) = \{ v_1, \ldots , v_n \}$, and we write
$\mu_k(i)$ for the $(i, i)$ entry of $A^k$, then the subgraph
centrality of node $v_i$, $C_{sg}(v_i)$ is defined by
\begin{eqnarray}
\label{eq:SubGCen} C_{sg}(v_i) = \sum_{k=0}^{\infty}
\frac{\mu_k(i)}{k!}.
\end{eqnarray}
The findings presented in \cite{Est05, Est05a} indicate that
$C_{sg}$ performs as well as node degree in predicting
essentiality.

Other concepts of centrality that have been proposed include {\it
flow betweenness centrality }\cite{FreBorWhi91}, {\it information
centrality }\cite{SteZel89}.  For completeness, we also note the
recent measure introduced in \cite{LatMar04} which ranks nodes
according to the effect their removal has on the efficiency of a
network in propagating information and the centrality measure
based on game theoretic concepts defined in \cite{Gometal03}.  We
shall not discuss these in detail here however as little work on
their biological relevance has been done to date.

\subsection{Alternative Approaches to Predicting Essentiality}

We shall now briefly discuss some other methodologies for
predicting gene or protein essentiality that have been proposed in
the last few years.

\underline{\bf Functional Classes and Essentiality}

In the Yeast Protein Database (YPD) \cite{Cosetal00} various
functional classes are defined to which the proteins in yeast can
be assigned.  Using the functional classification of proteins in
the Yeast Protein Database (YPD) \cite{Cosetal00}, the authors of
\cite{JeoOltBar03} studied the relationship between the functions
of a protein in the interaction network of yeast and its
likelihood to be essential.  They found that the probability of
essentiality varied significantly between the 43 different
functional classes considered. For instance, in one class
containing proteins that are required for DNA splicing, the
percentage of essential proteins was as high as $60\%$ while only
$4.9\%$ of the proteins in the class responsible for small
molecule transport were essential. This suggests that to predict
essentiality, the functional classification of proteins should be
taken into account.  However, the fact that many proteins are as
yet unclassified is a significant impediment to such an approach.

In the same paper, the nodes within each of the 43 functional
classes were ranked according to their degree and, within each
class, the degree of a protein was found to be a good indicator of
its likelihood to be essential. Genes were also ranked using the
standard deviation of their expression levels across a large
number of different yeast derivatives: each derivative
corresponding to one gene deletion.  Some connection between the
variability in the mRNA expression of a gene and its likelihood to
be essential was observed.  Specifically, genes whose expression
levels varied little were more likely to be essential.  It is
hypothesised in \cite{JeoOltBar03} that this may be due to
robustness mechanisms that maintain the expression levels of
essential genes close to a constant level, while those of less
important genes are subject to less stringent constraints, and
hence can be more variable.
%
%

\underline{\bf Damage in Metabolic and Protein Networks}

The concept of {\it damage} was recently defined for metabolic
networks in \cite{Lemetal04} and then later for protein
interaction networks in \cite{Schetal05}.  In the first of these
papers, metabolic networks were modelled as directed bi-partite
graphs \cite{Die00}.  Such a graph has two distinct sets of nodes:
one contains the metabolites while the nodes of the other set
represent the reactions catalysed by the enzymes of the
metabolism. Each such enzyme, $v$, is assigned a score $dg(v)$,
its {\it damage }, which characterises the topological effect of
deleting $v$ from the network.  Essentially, $dg(v)$ is the number
of metabolites that would no longer be produced if the enzyme $v$
and all the reactions catalysed by it were removed from the
network.  The following findings about the relationship of this
concept to essentiality were reported in \cite{Lemetal04}.
\begin{itemize}
\item[(i)] For each value of the damage, $D > 0$, let $f_{D}$ be
the fraction of enzymes, $v$ with $dg(v) = D$ which are essential.
An F-test indicated that there was a statistically significant
correlation between $D$ and $f_{D}$. \item[(ii)] The set of
enzymes $v$ for which $dg(v) \geq 5$ contains $9\%$ of all enzymes
and $50\%$ of the essential enzymes.
\end{itemize}
Based on their findings, the authors of \cite{Lemetal04} suggested
that enzymes with high damage are potential drug targets. However,
it should be noted that there exist several essential enzymes,
$v$, for which $dg(v)$ is quite low and that, conversely, there
are also non-essential enzymes with high damage scores.

More recently, in \cite{Schetal05} an analogous concept for
protein interaction networks was defined and applied to the yeast
protein interaction network.  The results of this paper indicate
that any correlation between damage and essentiality is very weak.
On the other hand, the authors of this paper found that if the set
of nodes disconnected from the network by the removal of a protein
$v$ contains an essential protein, then there is a high
probability of $v$ itself being essential.
%
%
Finally, we note another measure of importance in biological
networks which was recently described in \cite{PrzWigJur04}.  This
measure was based on the notion of {\it bottle-necks} within
networks and its relationship to essentiality was investigated in
this paper.

\subsection{Final Thoughts on Essentiality}

Finally, we shall discuss a number of issues with the various
approaches to predicting essentiality that have been described
throughout this section.

\begin{itemize}
\item[(i)] \underline{\bf Marginal Essentiality}

While our discussion has focussed on essentiality, a gene or
protein may be important to an organism without necessarily being
essential.  For instance, some sets of non-essential genes are
{\it synthetically lethal, } meaning that the simultaneous removal
of the genes in the set kills the organism while individual
deletions are non-fatal.  In the paper \cite{Yuetal04}, the less
restrictive concept of {\it marginal essentiality } and its
relationship to various topological measures was studied in the
protein interaction network of {\it S. cerevisiae}.  Here,
proteins were classified into five groups based on their marginal
essentiality: those with the lowest marginal essentiality scores
being assigned to group 1, and those with the highest assigned to
group 5.  The authors of \cite{Yuetal04} found that the average
degree and clustering coefficient of the nodes in a group
increases monotonically with the group number. For instance, the
average degree of those proteins assigned to Group 1 is about half
of that of the proteins in Group 4.  Moreover, defining a hub node
to be one in the first quartile of nodes ranked according to
degree, they found that less than $10\%$ of the proteins in Group
1 are hubs while more than $35\%$ of those in Group 5 are hubs.
The percentage again increased monotonically with the group
number.

\item[(ii)] \underline{\bf Fitness Effect and Evolutionary Rate}

In \cite{Fraetal02} it was reported that the degree of a protein
in the interaction network of yeast was positively correlated with
the {\it fitness effect }of deleting the gene that encodes the
protein.  Here, fitness effect measures the reduction in the
growth rate of the organism when the gene is deleted.  This
investigation was motivated by the question of whether the
importance of a gene or protein for an organism correlates with
the rate at which it evolves.  For more information, and varying
opinions on this topic, consult \cite{HahConWag04, Joretal02,
YanGuLi03, PalPapHur03, HirFra03, HirFra01}.

\item[(iii)] \underline{\bf Sensitivity to Data Errors}

The issue of sensitivity to data inaccuracy is of critical
importance for all of the techniques described here.  It was noted
in \cite{Schetal05} that the measure damage discussed above is
quite sensitive to false negative errors, in which a real
interaction between two nodes in a network has not been identified
due to experimental error.  Clearly, such sensitivity to data
noise has serious implications for the practical use of any of the
methods described here.  In particular, it is important to have a
thorough understanding of the effect of missing or inaccurate data
on the performance of centrality measures or other approaches to
predicting essentiality.  While there has been some research into
this fundamental issue recently \cite{CosVal03, ZemHle05,
BorCarKra05, Schetal05a}, more intensive quantitative and
theoretical studies are needed before we can reliably apply the
techniques discussed here to the problem of essentiality
prediction.  This issue is all the more important given that much
of the data available on bio-molecular networks contains large
numbers of false positive and false negative results
\cite{Coletal05, Meretal02}.

\item[(iv)] \underline{\bf Essentiality and Modules}

Finally for this section, we note the work of \cite{DezOltBar03}
on determining the essentiality and cellular function of modules
within the yeast PPI network.  The results of this paper indicate
that the essentiality (or non-essentiality) and functionality of
an overall complex is largely determined by a core set of proteins
within the complex.  Moreover, the essentiality of individual
proteins appears to depend on the importance of the modules in
which they lie.  This suggests that it may be more appropriate to
address the question of essentiality at the level of modules
rather than individual proteins or genes and motivates the problem
of extending centrality measures to deal with groups of nodes.
\end{itemize}

\subsection{Summarizing Comments}
\begin{itemize}
\item[(i)] In this section, we have discussed several measures of
the importance, or centrality, of the nodes in complex networks,
including degree centrality, betweenness centrality, closeness
centrality and eigenvector centrality.  We have described the
findings of several recent studies which have applied these
measures to datasets on protein-protein interaction and
transcriptional regulatory networks. \item[(ii)]  Most of the
studies discussed in the text indicate a link between the
centrality score of a gene or protein and its likelihood to be
essential for survival.  \item[(iii)]  There appears to be no
compelling evidence at the current time that the more complex
centrality measures described here perform any better as
indicators of essentiality than simple degree centrality.
\item[(iv)]  As with the identification of network structure
discussed in Section \ref{sec:models}, the impact of inaccurate
and incomplete data on the performance of centrality measures as
indicators of essentiality is of critical importance and needs to
be more fully investigated.
\end{itemize}

\section{Motifs and Functional Modules in Biological
Networks}\label{sec:motifs} The analysis methods discussed in the
previous section were concerned with identifying individually
important nodes within a network.  However, several recent studies
have revealed that bio-molecular networks are often modular in
nature, with groups of individual nodes collaborating to carry out
some specific biological function.  This has led researchers to
investigate more closely the hierarchical structure of real
interaction networks, and to provide biological explanations for
how the observed structure of such networks has emerged.

Recently, a loose hierarchical structure for bio-molecular
networks has been proposed in \cite{Babetal04, BarOlt04}.  The
lowest level in this hierarchy consists of individual nodes, which
are then organised into so-called {\it network motifs}.  Motifs
are small subgraphs that occur significantly more often in a
network than would be expected by chance.  These are in turn
grouped into larger modules of functionally related nodes before
finally, the modules are themselves connected to form the overall
network.  In this section, we shall discuss recent work on
identifying motifs within specific biological networks, and the
efforts of a number of researchers to use motifs to classify
networks into distinct families.  We shall also consider the
question of why motifs occur so frequently in real networks.
Towards the end of the section, we shall consider the problem of
identifying communities of functionally related nodes in
bio-molecular networks and discuss a number of algorithms that
have been proposed for this purpose.

\subsection{Identification of Network Motifs}

The concept of a network motif and a basic scheme for motif
detection were described in the paper \cite{Miletal02}.
Specifically, given a directed network $G$, the motifs in $G$ of
size $k$ are identified as follows:
\begin{itemize}
\item[(i)] For each possible subgraph, $S$ of size $k$, of $G$
count the number of occurrences, $N_S$, of $S$ in $G$. \item[(ii)]
Next randomly generate a large number of networks such that in
each random network:
\begin{itemize}
\item[(a)] Each node has the same in-degree and out-degree as in
the real network $G$; \item[(b)] Every subgraph of size $k-1$
occurs with the same frequency as in the real network $G$. Two
schemes for generating the random networks are described in
\cite{Miletal02} and its supporting material.
\end{itemize}
\item[(iii)]  A subgraph, $S$, is then said to be a motif of $G$
if it satisfies the following three conditions:
\begin{itemize}
\item[(a)] The probability of $S$ occurring in a random network
more often than $N_S$ times is less than some prescribed value $P$
(in \cite{Miletal02} $P$ is taken to be $.01$); \item[(b)]  There
are at least four distinct occurrences of $S$ in the network $G$;
\item[(c)]  The actual number of occurrences of $S$ in $G$ is
significantly larger than the average number of occurrences of $S$
in the randomly generated networks, denoted $\langle N_S^{rand}
\rangle$; formally, $N_S - \langle N_S^{rand} \rangle \,
> 0.1 \langle N_S^{rand} \rangle$.
\end{itemize}
\end{itemize}

\underline{\bf Comments}
\begin{itemize}
\item[(i)] The scheme described above can, and has been
\cite{WucOltBar03}, easily adapted to detect motifs in undirected
networks such as protein interaction networks. \item[(ii)]  The
identification of motifs within large complex networks is
computationally intensive and, to the best of the authors'
knowledge, standard methods are only feasible for motifs
containing less than 7 or 8 nodes. \item[(iii)] In
\cite{Zivetal05} a systematic method of defining network measures
or ``scalars'' which are related to subgraphs and can be used to
detect motifs was introduced.  The techniques of this paper
address some of the issues with standard motif detection
algorithms but the precise relationship between ``scalars'' and
subgraphs is not straightforward.
\end{itemize}

Using the scheme described above, small motifs have been
identified in a number of real biological networks.  In
particular, the transcriptional regulatory networks of {\it E.
coli} and {\it S. cerevisiae} have been found to have one
three-node motif and one four-node motif. These are the so-called
{\it feed-forward }motif and {\it bi-fan }motif, shown in Figure
\ref{fig:MotifTran} below.
\begin{figure}[H]
             \begin{center}
                \psfrag{Ff}[ct][ct][0.9][0]{Feed-forward loop}
                \psfrag{Bf}[ct][ct][0.9][0]{Bi-Fan}
                \psfrag{w}{$w$}
                \psfrag{x}{$x$}
                \psfrag{y}{$y$}
                \psfrag{z}{$z$}
               \includegraphics[width=.6\columnwidth]{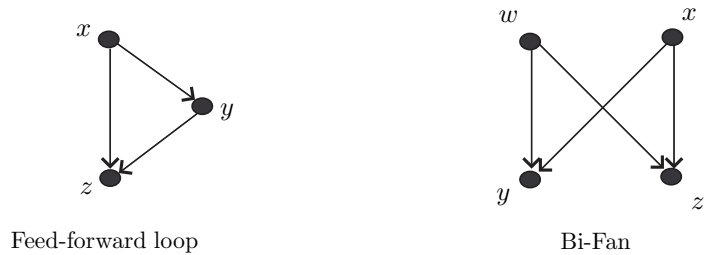}
               \caption{Feed-forward and Bi-Fan motifs of transcriptional networks}
               \label{fig:MotifTran}
             \end{center}
\end{figure}
The feed-forward and bi-fan patterns are also motifs of the
neuronal network of the nematode {\it C. elegans}.  This network
has an additional four-node motif known as the bi-parallel motif.
Other common motifs which have been detected in food webs,
electronic circuits and the World-Wide-Web include the {\it
three-chain }, {\it three and four-node feedback loops } and the
{\it fully-connected triad } shown below.  Note that the network
motifs of the transcriptional network of yeast have also been
investigated in the paper \cite{Leeetal02}, where the motifs
identified have also been related to specific information
processing tasks.
\begin{figure}[H]
             \begin{center}
                \psfrag{TC}[ct][ct][0.9][0]{Three-chain}
                \psfrag{TNFF}[ct][ct][0.9][0]{Three-node feedback}
                \psfrag{BP}[ct][ct][0.9][0]{Bi-parallel}
                \psfrag{FNFF}[ct][ct][0.9][0]{Four-node feedback}
                \psfrag{w}{$w$}
                \psfrag{x}{$x$}
                \psfrag{y}{$y$}
                \psfrag{z}{$z$}
               \includegraphics[width=.8\columnwidth]{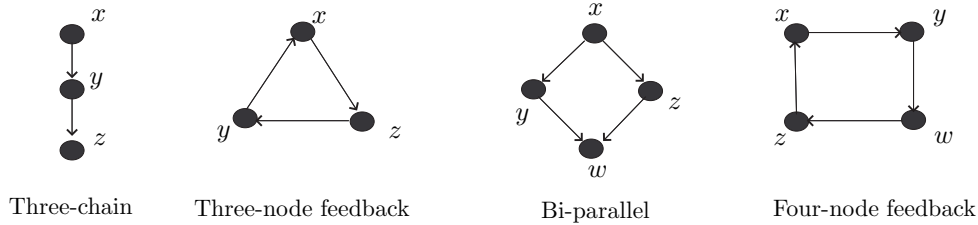}
               \caption{Common motifs in real networks}
               \label{fig:MotifGen}
             \end{center}
\end{figure}

Before proceeding, a number of facts about the findings reported
in \cite{Miletal02} are worth noting.  The feed-forward loop and
bi-fan motifs have been found in transcriptional regulatory
networks and neuronal networks, both of which involve some form of
information processing.  Also, the motifs found in the food-webs
studied are distinct from those found in transcriptional
regulatory networks and the WWW, while electronic circuits with
distinct functions tend to have different sets of motifs. These
observations have led some authors to suggest that there is a
connection between a network's motifs and its function, and hence,
that complex networks may be classified into distinct functional
families based on their typical motifs.  For instance, given that
information processing is fundamental to both neuronal and
transcriptional networks, it is reasonable to suggest that
feed-forward loops and bi-fans occur often in such networks
because of their suitability for information processing tasks.  On
the other hand, there is no overlap between the motifs observed in
transcriptional networks and those of the functionally unrelated
food-web networks.

Finally for this subsection, we note that the transcriptional
network of {\it E. coli} has been investigated in more detail in
\cite{Shenetal02} and several additional motifs have been
identified: {\it single input modules (SIMs) }; {\it dense
overlapping regulons } and {\it negative autoregulatory units}.

\subsection{Dynamical Properties of Motifs}

In the recent past, there have been several theoretical and
experimental studies carried out on the dynamical properties of
specific motifs and on clarifying the relationship between motifs
and functionality. For example, in \cite{Shenetal02}, it was
demonstrated that the feed-forward loop motif can provide a
mechanism of filtering out transient or fluctuating input signals.
This motif structure also responds to persistent activation with a
slight delay and shuts down rapidly once the activating signal is
removed.  Circuits of this type are said to act as {\it
sign-sensitive delays}.  The authors of \cite{Shenetal02}
presented a simple mathematical model to describe the action of
the feed-forward loop in transcriptional regulatory circuits and
then studied the behaviour of this model under the assumption that
the circuit was {\it coherent }in the following sense.  Each
regulatory interaction is assigned a positive or negative sign
depending on whether it is excitatory or inhibitory.  The circuit
is coherent if the indirect and direct paths have the same sign,
and incoherent otherwise.  Under the assumption that all
regulations are excitatory, it was shown numerically in
\cite{Shenetal02} that the FFL motif does indeed act as a
sign-sensitive delay element.

A more complete mathematical analysis of the kinetic behaviour of
the FFL motif was presented in the paper \cite{ManAlo03}, where
the response times of all of the different possible configurations
of the FFL were studied.
Note however that coherent configurations seem to occur far more
frequently than incoherent configurations in real systems such as
the transcriptional network of {\it E. coli} \cite{ManZasAlo03}.
Also, in \cite{HayJay05}, a more detailed model of the coherent
FFL circuit was described and analysed.  Here, the robustness of
the model's behaviour with respect to variations in parameter
values and external perturbations was investigated. For instance,
the sign-sensitive delay action was found to be quite sensitive to
variations in the model's parameters and, while the circuit is
quite robust with respect to the size of external perturbations,
the duration of the perturbation in comparison to the internal
time-scales of the circuit appears to be critical.

In addition to the theoretical investigations described above, the
kinetics of the coherent FFL motif have been studied
experimentally in \cite{ManZasAlo03}.  Specifically, the authors
of this paper analysed the {\it l-arabinose }utilization circuit
in {\it E. coli} and confirmed that, in this case, the coherent
FFL circuit functions as a sign-sensitive delay element that
filters out transient activation signals from a fluctuating
environment.

Before finishing our discussion of this topic, we should note a
number of other theoretical and experimental investigations of the
dynamical properties of network motifs.  The negative
autoregulatory circuit consisting of a transcription factor that
down-regulates its own transcription was studied in
\cite{RosEloAlo02}, where the response times of a simple
transcriptional unit (without autoregulation) and a negative
autoregulatory circuit were compared.
Here, it was shown theoretically that the response-time of the
autoregulatory circuit is shorter than that of the simple
transcriptional circuit, with the same steady state. In fact, for
very strong auto-repression, the response-time of the
auto-regulatory circuit is only one fifth of that of simple
transcription.  It has also been demonstrated experimentally in
the same paper that while a transcriptional circuit without
autoregulation has a response-time of approximately one
cell-cycle, the response-time for a circuit with negative
auto-regulation is about one-fifth of a cell cycle. Finally, we
also note the recent work on the kinetics of the single-input
module (SIM) motif in {\it E. coli} \cite{Shenetal02} and the
p53-Mdm2 feedback loop \cite{Lahetal04}.

\subsection{Evolutionary Conservation, Extensions and Final Thoughts on Motifs}

\underline{\bf Motifs and Evolutionary Conservation}

The work discussed in the last subsection was concerned with
investigating the dynamical properties and biological function of
a number of common motifs. The biological significance of motifs
has been considered from a slightly different point of view in
\cite{WucOltBar03} where the extent to which motifs in the protein
interaction network of yeast are evolutionarily conserved was
studied. Specifically, 678 proteins in the yeast PPI network were
identified which have orthologs \footnote{Orthologs are genes with
a common ancestor.} in each of five higher organisms, and for each
2, 3, 4 and 5 node motif, the percentage of motifs which were
completely conserved across all of the 5 higher organisms was
determined.  A sub-graph is completely conserved if all of the
proteins in it have orthologs in each of the higher organisms. For
the yeast PPI network, motifs which have a higher number of nodes
and are more densely interconnected also have a higher rate of
conservation. For instance, the completely connected five-node
motif has the highest rate of conservation of all motifs with
between 2 and 5 nodes.

To validate these findings, the same number of orthologs was
positioned randomly on the network and the percentages of
completely conserved motifs were again calculated. In this case,
the rates of conservation were considerably lower, and moreover,
the rate of conservation decreased with increasing motif size, in
contrast to what was observed for the real orthologs.  In
particular, for the completely connected five-node motif, the
natural rate of conservation was found to be $47.24 \%$ while the
random conservation rate was as low as $.02 \%$. Furthermore,
larger, more tightly connected and conserved motifs were found to
be more functionally homogeneous.  In fact, for a significant
number of these, all of the proteins in the complex belonged to at
least one common functional class.

Note also that in \cite{morenovega04} a correlation between the
natural rate of conservation of motifs in the yeast PPI network
and the suitability of the motif structure for synchronization of
interconnected Kuramoto oscillators was reported.  We shall have
more to say about the question of synchronization later in the
article.

\underline{\bf Extensions of the Motif Concept}

In \cite{Miletal04}, the significance profile (SP) was proposed as
a means of classifying networks.  Given a network, $G$, for each
possible subgraph, $S$, the number of occurrences of $S$ in a real
network $G$ is calculated and compared to the average number of
occurrences of $S$ in an ensemble of random networks with the same
degree profile as $G$. The $Z$-score for each such subgraph is
then calculated as
\begin{eqnarray}
Z_S = \frac{N_S - \langle N^{rand}_S \rangle}{std(N^{rand}_S)}
\end{eqnarray}
where $N_S$, $\langle N^{rand}_S \rangle$ and $std(N^{rand}_S)$
denote the number of occurrences of $S$ in $G$, and the mean and
standard deviation of the number of occurrences of $S$ in the
ensemble of random networks respectively.  The vector of
$Z$-scores for subgraphs of a fixed size is then normalized to
give the {\it significance profile }vector.
\begin{eqnarray}
SP_S = \frac{Z_S}{(\sum_S Z_S^2)^{1/2}}.
\end{eqnarray}
Significance profiles for subgraphs of sizes three and four are
calculated in \cite{Miletal04} for a number of real biological
networks.  While this method has been proposed as a means of
identifying different classes of complex networks, it should be
noted that some networks with similar SP vectors for three-node
subgraphs have distinct four-node SPs.  As mentioned in
\cite{Miletal04}, this means that higher order SPs are needed if
this technique is to be used effectively to classify networks.
Also it is not clear at the moment how to determine the maximal
subgraph size required to correctly distinguish network classes
using this technique.

Another possible extension of the motif concept was recently
suggested in \cite{Kasetal04}.  Here, so-called {\it topological
generalizations }of subgraphs and motifs were introduced based on
duplicating certain nodes within the subgraph.  Several
significant motif generalizations within the transcriptional
regulatory networks of {\it E. coli} and {\it S. cerevisiae} were
identified and possible functions for the observed generalizations
were also proposed and investigated on simple mathematical models
of transcriptional regulation and neuronal networks.  While most
of our discussion has focussed on transcriptional networks or
protein interaction networks in isolation, this distinction is
somewhat artificial, and ultimately the methods described here
will need to be extended to more integrated cellular networks.  In
this context, the work of \cite{Yegetal04} on identifying motifs
within a more complete cellular network, which takes into account
both transcriptional interactions and direct protein-protein
interactions, and the study of motifs within an integrated network
involving five different interaction types in \cite{Zhaetal05}
should be noted.

\underline{\bf Some Final Thoughts on Motifs}

Before drawing our discussion of network motifs to a close, we
note a number of motivations for the study of network motifs as
well as some caveats that should be kept in mind.
\begin{itemize}
\item[(i)] Studying the motifs of a complex biological network can
provide useful insights into the both the structure and function
of the network.  For instance, once we have identified a network's
motifs, analysis such as that described above on the dynamical
properties of the FFL motif can help us to determine the key
functional roles of the network. Motifs can also be used to help
develop more complete models for the evolution of bio-molecular
networks than those discussed in Section \ref{sec:models}.
\item[(ii)] As mentioned above, motifs and extensions such as the
significance profile could be used to identify distinct categories
of complex networks. However, as noted in \cite{Miletal04},
networks with the same motif profile for three-node subgraphs can
have different four-node or higher order motifs and this casts
some doubt on how effective these methods are likely to prove as a
means of classifying networks. Moreover, the identification of
higher order motifs is likely to be very costly from a
computational point of view.  \item[(iii)] A knowledge of the
motifs of a network is a necessary step in unravelling its
hierarchical structure and can help in identifying modules which
can then be used to simplify the network's analysis. \item[(iv)]
The precise biological significance of the various network motifs
which we have discussed is still unclear and it should be noted
that while motifs are {\it statistically }significant subgraphs,
there may be other subgraphs within a network, occurring in
smaller numbers, that are biologically important.  This issue has
been debated in \cite{Ranetal04, Miletal04a}, and in
\cite{ConWag03} two biological reasons for the emergence of motifs
have been considered: gene duplication and convergent evolution.
The findings described in \cite{ConWag03} indicate that the motifs
in the transcriptional regulatory networks of {\it S. cerevisiae}
and {\it E. coli} have not emerged due to gene duplication, which
the authors argue, provides evidence for claims that network
motifs have emerged as the result of some mechanism of natural
selection. Further evidence that motifs have emerged as a result
of some biological optimization was recently presented in
\cite{ItzAlo05}. Here, the motif patterns of {\it geometric
networks, }where links are formed based on the spatial proximity
of nodes, were studied analytically.  The results of this paper
show that simple geometric constraints alone are not sufficient to
account for the motifs observed in biological and social networks.
The authors of this paper argued that this indicates that
additional, possibly biological, factors have played a role in
determining the emergence of motifs in such networks.
\end{itemize}

\subsection{Community Structure and Functional Modules in Biological Networks}

In the introductory remarks to this section, we outlined a loose
hierarchical structure for biological networks, in which the next
level above that of motifs was that of modules or communities of
functionally related nodes.  Recent research has indicated that
functionally coherent families of genes and proteins can be
determined from the topology of interaction networks
\cite{YooOltBar04}. Hence, the development of reliable methods for
identifying such {\it functional modules }would have significant
implications for the problem of assigning functions to unannotated
proteins or genes. This issue is of considerable importance given
that the biological function of many of the genes and proteins
within even simple organisms such as yeast are still unknown.  In
this subsection, we shall discuss a number of algorithms and
techniques that have been developed for the identification of
modules and community structure within complex networks.  To begin
with, we shall review those methods that have been developed
specifically to detect functional families and hierarchical
structure in bio-molecular networks.  We shall also discuss some
techniques that have been proposed specifically for the problem of
protein function assignment.

\underline{\bf Network Hierarchy and Motif Clusters}

The authors of \cite{Dobetal04} studied how the FFL and bi-fan
motifs in the {\it E. coli} transcriptional regulatory network are
integrated into the overall network structure.  The findings of
this paper suggest a hierarchical organization of this network,
where motifs are first aggregated into larger {\it motif clusters,
}with each cluster primarily consisting of the same motif type.
These clusters are then further combined into so-called {\it
super-clusters }which form the core of the overall network.  For
instance, all but one of the identified feed-forward loops (FFLs)
in the network were contained in six FFL clusters, and similarly,
all but one of the bi-fan motifs were contained in two bi-fan
clusters.  Moreover, these motif clusters combined to form one
large super-cluster containing all but one feed-forward loop and
one bi-fan motif.

Another approach to investigating the hierarchical and modular
structure of the transcriptional network of {\it E. coli} was
described in \cite{MaBueZen04}.  Here, five different regulatory
levels were identified, such that each node is either
self-regulatory or else can only regulate nodes at lower levels.
Based on this hierarchical decomposition of the network, a scheme
for identifying modules of functionally related genes was
described which appears to work quite well in identifying sets of
genes with similar functionality.  The authors of this paper also
found that many of the FFL and bi-fan motifs in this network
contained genes responsible for regulating modules with diverse
function.  They argue that this fact is not in agreement with the
view that motifs themselves form the basic building blocks of
functional modules, as, for instance, it shows that the same
feed-forward loop can be involved in the regulation of numerous
different modules.

\underline{\bf Graph Theoretical Approaches to Identifying
Functional Modules}

A graph clustering algorithm for identifying families of related
nodes in networks was described in \cite{EnDonOuz02}, where the
problem of how to cluster proteins in large databases into
families based on sequence similarity was considered.  The first
step in this algorithm was to assign sequence similarity scores to
each pair of proteins using an algorithm such as BLAST.  A
weighted graph was then constructed, whose nodes are proteins and
where the weight of an edge between two nodes is the similarity
score calculated in the previous step.  The TRIBE-MCL algorithm
for detecting communities of related nodes within this graph was
then described.  This technique is based on {\it Markov chain
clustering }, and identifies communities through iterating two
different mathematical operations of {\it inflation }and {\it
expansion }.  The core concept behind this method is that families
of related nodes are densely interconnected and hence there should
be more ``long'' paths between pairs of nodes belonging to the
same family than between pairs of nodes belonging to distinct
families.  Subsequently, in \cite{PerEnOuz04} this algorithm was
used to identify functionally related families in the protein
interaction network of {\it S. cerevisiae}.  In fact, the
algorithm was applied to the {\it line-graph }$L(G)$, where the
nodes of $L(G)$ are the edges of $G$ and two nodes in $L(G)$ are
connected if the corresponding edges in $G$ are incident on a
common node in $G$. Three separate schemes of protein function
classification were then used to validate the modules identified
with this algorithm, and the coherence of functional assignment
within these modules was significantly higher than that obtained
for random networks obtained by shuffling protein identifiers
between modules.  This together with further analysis indicated
that the identified modules did represent functional families
within the network.

Further approaches to the determination of functional modules
within biological networks have been described in
\cite{PrzWigJur04, Segetal03}.  The technique in
\cite{PrzWigJur04} relies on searching for {\it highly connected
subgraphs (HCS)} where a HCS of a graph $G$ is a subgraph $S$ for
which at least half of the nodes of $S$ must be removed in order
to disconnect it.  On the other hand, in \cite{Segetal03,
Segetal03a} a procedure is described which identifies modules of
related genes in the transcriptional regulatory network of yeast
as well as the regulators of each such module. Other approaches to
determining functional modules within transcriptional networks
have been described in \cite{Baretal03, Ihmetal02}.  The
techniques described in these papers are not based on a graph
theoretical analysis of network topology however; in fact, they
rely on analysing gene expression data across different
experimental conditions and determining sets of genes which are
regulated by common transcription factors.

\underline{\bf Predicting Protein or Gene Function from Network
Structure}

Several direct approaches to assigning functions to unannotated
proteins have also been proposed recently.  The simplest of these
is the so-called {\it majority rule }which works in the following
way \cite{SchUetFie00, Nabetal05}.  Given a classification scheme
with an associated set of functions,
\[ \mathcal{F} = \{f_s: 1 \leq s \leq M \},\]an interaction network,
$G$, and an unannotated protein $i$ in $G$, each function, $f_s
\in \mathcal{F}$, is assigned a score which is simply the number
of times $f_s$ occurs among the annotated neighbours of $i$.  The
functions with the highest scores are then identified as the most
likely functions for the protein $i$.  A simple extension of this
concept which takes into account nodes other than the immediate
neighbours of the unannotated protein was presented in
\cite{Hisetal01}.  It should be noted that this approach has the
major drawback of relying entirely on the functions of previously
annotated proteins, while it can often happen that none of the
neighbours of a protein of unknown function have been annotated.

Two more sophisticated approaches to protein function prediction
that avoid the above mentioned difficulty were described in
\cite{Vazetal03a, Karetal04}. Essentially, these algorithms assign
functions to the proteins in an interaction network so as to
minimize the number of pairs of interacting proteins with
different functional assignments.  A key aspect of these
approaches is that the optimal global assignment of protein
function is not unique.  In practice, a number of different
optimal solutions are determined, and the frequency with which a
given function $f_s$ is assigned to a protein $i$ is interpreted
as the probability of the protein having that function.

The work presented in two other recent papers is also worth noting
in the present context.  Firstly, in \cite{Nabetal05}, the {\it
functional flow }algorithm was described.  The core idea of this
method is to consider annotated proteins within the network as
reservoirs or sources of flow for the functions assigned to them.
Each such function then ``flows'' through the network according to
a specified set of rules and the amount of each function at a node
when the iterations finish is used to determine the most likely
functions for that node.  On the other hand, the technique
described in \cite{SamLia03} is based on the hypothesis that pairs
of proteins with a high number of common interaction partners are
more likely to share common functions.  Formally, for a pair of
proteins $i$, $j$, of degrees $n_1$, $n_2$ respectively, with $m$
common interaction partners, the probability $p(i, j, m)$ of them
having $m$ common partners if links were distributed randomly is
calculated.  This method was applied to the protein interaction
network of {\it S. cerevisiae} and, of the 100 pairs of proteins
with the lowest value of $p(i, j, m)$, over $95 \%$ of them
consisted of proteins with similar function.  The authors also
described how to use these basic ideas to identify modules within
an interaction network and validated the method on the yeast
interaction data.  A related probabilistic approach to using
interaction network topology to predict protein function has also
been presented in \cite{LetKas03}.

In the recent paper \cite{Segetal05}, the PRISM algorithm for
identifying modules of functionally related genes based on
analysing {\it epistatic} interactions was presented.
\footnote{Epistatic networks describe the interactions through
which different genetic mutations either aggravate or buffer each
other's effects on an organism.} The core idea behind this
algorithm is that genes belonging to one functional module should
interact with genes in another module in a similar fashion.  Using
this algorithm, it was possible to group genes with similar
functional annotation into the same module even in the absence of
a direct interaction between them.  Finally, we note that in
\cite{Buetal03}, a technique for identifying {\it quasi-cliques }
in protein interaction networks based on the eigenvectors of the
network's adjacency matrix was described and applied to the yeast
interaction network.  Most of the quasi-cliques identified in this
way were found to have homogeneous functional annotation in the
MIPS database suggesting that this technique could be useful in
assigning function to unannotated proteins.

\underline{\bf Module Identification in General Networks}

The problem of identifying communities and cohesive modules is
also of importance in a number of other application domains
including the analysis of social networks, communication networks,
and power networks. Before concluding this section, in the
interest of completeness, we briefly note some techniques that
have been developed recently for this general problem which could
also be used to identify functional modules within bio-molecular
networks; see \cite{New04b} for an overview of recent and
traditional approaches to the problem of identifying community
structure in networks.  In \cite{GirNew02, NewGir04, Radetal04,
ForLatMar04} the notions of betweenness and information centrality
were first extended to apply to edges rather than nodes. Then,
based on the observation that edges connecting distinct
communities will typically have higher betweenness and information
centrality scores than edges within communities, divisive
algorithms for identifying communities were described. The basic
principle of these algorithms has been adapted to determine the
hierarchical and modular structure of metabolic networks in
\cite{HolHusJeo03}.  More recently, algorithms based on
edge-betweenness have been applied to datasets on protein
interaction networks in yeast and humans in \cite{DunDudSan05}.
Here, the ability of these clustering algorithms to identify
families of functionally related proteins was studied.  The
robustness of these approaches to false positive errors in the
datasets was also investigated.

In contrast to the divisive approaches discussed above, the
techniques proposed in \cite{New04, ClaNewMoo04} work in an
agglomerative rather than divisive fashion.  Here, a function that
measures the {\it modularity }of a proposed division of a network
into communities was defined and algorithms to optimize this
function appropriately were presented. A different,
information-theoretic measure of modularity which applies directly
to a network rather than a specific partition of the network has
recently been proposed in \cite{ZivMidWig05}. Algorithms for
splitting a network into modules were also described in the same
paper and their effectiveness was tested on real and synthetic
network data, with promising results.  Note also the approaches
based on analysis of the spectrum of the Laplacian matrix of the
network described in \cite{Capetal05, DonMun04}.

\subsection{Summarizing Comments}

\begin{itemize}
\item[(i)] In many real biological, and technological, networks,
certain small subgraphs occur far more frequently than would be
expected for randomly wired networks with the same degree
distribution. Such subgraphs are known as motifs. \item[(ii)]
Experimental observations have indicated that networks with
similar function tend to have similar sets of motifs.  For
instance, feed-forward loops are very common in both neuronal and
transcriptional regulatory networks; both of which are involved in
the processing of biological information.  This has led
researchers to consider a network's motifs as being characteristic
of the network in some sense. \item[(iii)]  While the precise
biological significance of motifs is still not completely
understood, several recent studies on the dynamical properties of
simple motifs have provided some insights into this question.  In
particular, the dynamics of the FFL motif and the auto-regulatory
motif in transcriptional networks have been studied and linked to
biological function. \item[(iv)] Graph theoretical techniques have
been used to determine the role of proteins or genes whose
function is currently unknown.  Several such techniques have been
described in the text.  General algorithms for identifying
communities in complex networks can also be applied to protein
interaction networks to identify modules of functionally
homogeneous proteins.
\end{itemize}

\section{Synchronization}
So far, our discussion of complex biological networks has largely
focussed on their \emph{structural} properties.  We have discussed
some of the key topological parameters for bio-molecular networks,
as well as numerical techniques for identifying the most important
nodes within such networks and for elucidating their hierarchical
structure.  As a next step, we shall consider some aspects of
network \emph{function} and network \emph{dynamics}, with
particular emphasis on the relationship between dynamic behaviour
and network topology.  Specifically, in this section, we shall
review the results of recent work on the connection between
synchronizability|that is, fitness for synchronization| and
network properties such as characteristic motifs, average
node-degree, betweenness centrality and degree distribution.

The outline of the section is as follows. We start off with a
general introduction to synchronization phenomena. We then
consider some aspects of mathematical modelling, focusing on the
Kuramoto model of coupled oscillators, and discuss some common
measures of synchrony. Following this, we review results on
synchronizability. Lastly, we discuss the role of synchronization
in the brain with particular emphasis on the connection between
abnormal synchrony and neurological disorders. We close with some
summarizing comments.

\subsection{Biological Oscillators and Synchrony}

Consider a graph wherein each node represents a dynamical process
and each edge an interaction between two processes. This simple
construct makes for an elegant description of many biological
systems. Consider, for example, the group of pacemaker cells that
make up the Sinoatrial Node \cite{Gla01} in the heart. For this
system we can construct a graph, such that each node represents a
pacemaker cell and each edge an interaction between two such
cells. The topology of this graph determines, to a large degree,
the system's overall behaviour. Indeed, abnormalities in the way
these cells are wired up can have a more significant impact on the
functioning of the Node than defects in individual cells. To
appreciate how important these interactions are, consider that in
isolation, each pacemaker cell oscillates at its own distinct
frequency; yet when put together, these same cells coordinate
their action in such a way as to generate a single impulse exactly
once during each cardiac cycle \cite{keener98}. This is an
instance of a phenomenon known as synchronization, or more
precisely, frequency synchronization. Roughly speaking,
synchronization is the process through which the output of a
system aligns itself with that of another system or group of
systems. A special case of this is the synchronization of
oscillators, to which we shall confine ourselves in this survey.

Oscillators can lock to both internal and external stimuli. The
locking to external stimuli is generally called
\emph{entrainment}. Examples of entrainment are commonplace in
human physiology. Many of the fundamental rhythms in our body, for
instance, are entrained by the light-dark cycle \cite{Gla01}. More
specifically, neural circuits have been found to support
entrainment, particularly in the gamma frequency range (50-100
Hz)~\cite{odonnell02}.

Synchronization is a population effect in the sense that it
emerges in complex systems comprising a large number of identical
or nearly identical components. In the natural world,
synchronization manifests itself across many different levels of
organization, from groups of organisms (the synchronous flashing
of fireflies \cite{buck88}) down to groups of cells (the pacemaker
cells in the example of the Sinoatrial Node). Less well known,
perhaps, is its implication in discussions on the binding problem,
one of the central problems in the philosophy of mind.
Specifically, in this context, synchronization has been put
forward as a mechanism to explain how information, distributed
across the brain, might be integrated to make coherent perception
possible~\cite{engel01}. Given the variety of applications, the
importance of understanding the principles of synchronization is
clear. One way to gain such understanding is to try to reproduce
this phenomenon \emph{in silico}, using a simple mathematical
model of coupled oscillators. In the next subsection we review
some aspects of the Kuramoto model, which has been the principal
model used for the study of synchronization phenomena over the
past thirty years.

\subsection{A Model of Synchronization} Over the years, a great deal
of interest has been expressed in the physics and mathematics of
synchronization. One of the first to present a detailed
mathematical treatment of the subject was Arthur Winfree. His 1967
paper \cite{winfree67} laid the basis for the work of Kuramoto and
others, who helped develop it into a mature mathematical theory
with applications in different fields \cite{strogatz03}.

In this review we shall focus on a model of synchronization,
introduced and popularized by Kuramoto~
\cite{kuramoto75,kuramoto87,acebron05}:
\begin{eqnarray}\label{eq:kuramotomodel}
    \dot{\theta}_i & = & \omega_i + \frac{K}{N}\sum_{j=1}^N
    \sin(\theta_i-\theta_j)
\end{eqnarray}
Here $\theta_i$ and $\omega_i$ respectively denote the phase and
intrinsic frequency of oscillator $i$; $K$ is the coupling
strength, and $N$ is the number of oscillators. This setting
assumes undirected all-to-all coupling, meaning that the
underlying graph is complete \cite{Die00}

Kuramoto's model describes a mechanism of self-organization in a
population of coupled oscillators. The model is thought to reflect
some aspects of coordinated behaviour in natural systems. Studies
indicate that the emergence of synchronization in the model is
robust with respect to variations in the interconnection
structure, albeit that the transition dynamics generally depend
upon the details of the underlying topology. We shall discuss this
dependence in more detail shortly.

A qualitative description of the behaviour of the
system~(\ref{eq:kuramotomodel}) is as follows (see
\cite{strogatz00}). When the interactions are weak, i.e.~$K$ is
small, the system is in an incoherent state, in which the
distribution of the phases $\{\theta_i\}$ is roughly uniform. In
this state, each oscillator tends to oscillate at its own
intrinsic frequency, $\omega_i$. When the level of interaction is
gradually increased, clusters of oscillators will emerge,
oscillating at a common frequency and (sometimes) phase. When the
coupling is still further increased, more and more oscillators
will join in, leading eventually to a state of full
synchronization in which all oscillators are oscillating as one.
Note that, strictly speaking, full synchronization is only
possible when all the oscillators are identical, i.e.~when
$\omega_i=\omega_j$ for all $i,j$. The transition from a
completely incoherent to a completely coherent state is typically
steep, and associated with it is some critical coupling strength
$K_c$, which marks the start of this transition.

The analysis of the Kuramoto model has a long and rich history,
and while a full understanding of its dynamics is still lacking, a
number of important results have been obtained in recent years.
Among them a proof of the instability of the unsynchronized state
for large coupling strengths, and formulae for the critical
coupling, and the steady state coherence. Most of these results
are only strictly valid in the thermodynamic limit when
$N\rightarrow \infty$, though some results are available for large
but finite populations \cite{jadbabaie04}. For an extensive review
of results related to the Kuramoto model and its applications, the
reader may consult \cite{acebron05}.

The Kuramoto model has found application in many areas, including
neuroscience, physics, engineering and biology
\cite{pikovsky01,acebron05}. This wide applicability appears to be
both a strength and a weakness in the sense that, as the model
captures the essence of synchronization, it necessarily lacks the
specificity to fully describe any one phenomenon in particular. We
shall come back to this when we discuss the role of
synchronization in the brain.

\subsection{Types and Measures of Synchrony}
In a system of coupled oscillators such as
(\ref{eq:kuramotomodel}), the emergence of synchronization is easy
to detect and quantify. In experiment, this is not quite as easy.
The fundamental problem is to extract from the complex time series
that are your data information about phase and frequency. This is
a non-trivial problem as the underlying processes are typically
non-stationary and, in a strict sense, non-periodic.

Before trying to detect synchronization proper, there are a few
other things one can do. For instance, to test for statistical
dependence between two time series, one could compute the spectral
covariance or \emph{coherence} \cite{peebles00}. In
\cite{roelfsema97,seidenbecher03} this technique was used to
quantify task-specific interactions in the brain. In recent years,
it has been suggested that this measure would lack the sensitivity
required to detect subtler forms of synchrony, such as phase
synchrony, as it would not separate out effects of amplitude and
phase.

Other measures of synchrony include {phase coherence}
\cite{spencer03,lachaux99}, {entropy}, and mutual information
\cite{hurtado04,ioannides04}. These latter measures are
particularly popular among experimentalists, who seek to
establish, for instance, whether or not a particular phase
relationship exists between a given set of experimental variables.
The application of these measures is limited by the fact that, in
a typical experiment, phase information is not directly
accessible, but needs to be extracted from the recorded time
series using specialized algorithms. This is a nontrivial problem
as the time series (e.g.~EEG recordings) are generally
non-periodic, and hence standard notions of phase do not apply.
Fortunately, there exist alternative notions of phase that do
generalize to non-periodic signals. Based on these notions,
computational techniques have been developed that are capable of
extracting phase information from arbitrary time series
\cite{pikovsky01}. These techniques have been successfully applied
to the analysis of brain data
\cite{lachaux99,hurtado04,spencer03}, revealing interesting
patterns of synchrony.

Another factor that might complicate the application of these
measures in practice, is the lacks of statistics. If prior
information about the data were available one could use that to
specify what degree of coherence should be considered
statistically significant. But in an experimental setting, such
information is typically not available. One way to overcome this
problem is to use schemes which generate ensembles of surrogate
data that are in some sense statistically similar to the original
time series \cite{hurtado04}. An early example of an application
of this approach can be found in \cite{lachaux99}.

For the system of coupled oscillators (\ref{eq:kuramotomodel}),
the standard measure of synchrony is the order parameter,
typically but not exclusively defined as
\cite{mcgraw05,hong04,strogatz00}:
\begin{eqnarray}
    r(t) & = &
    \left|\frac{1}{N}\sum_{j=1}^Ne^{i\theta_j(t)}\right|,
\end{eqnarray}
where, as before, $N$ denotes the number of oscillators in the
network, and $\theta_j(t)$ the instantaneous phase of oscillator
$j$. Geometrically, the value of the order parameter indicates how
well a given set of unit vectors are aligned with respect to one
another (with $1$ indicating perfect alignment). A slightly more
general definition is adopted in ~\cite{restrepo05}, incorporating
the adjacency matrix to account for the network's local structure.
Much the same measure is used again in \cite{ichinomiya04}.

\subsection{Synchronizability}
Recent studies have indicated that particular network properties,
such as the average clustering coefficient and betweenness
centrality, among others, have a major impact on the dynamics of a
network \cite{mcgraw05,ichinomiya04,hong04,gong05}. Here we shall
review those results that relate specifically to synchronization.

 \underline{\bf Kuramoto Oscillators on Random
Graphs}

The study of systems of coupled oscillators has recently been
extended from dealing exclusively with networks with all-to-all
coupling to include networks with local connectivity, such as
lattices and, indeed, random networks (particularly scale-free and
small-world networks). It has become clear that these complex
networks differ from their regular (random) counterparts in many
ways, and not least in terms of their dynamic properties. Great
interest has been expressed in the question as to what extent the
topology of a network determines the behaviour of the same; or
more in particular for a system of coupled oscillators: to what
extent the topology impacts the transition behaviour.

As regards the latter question, in~\cite{restrepo05} the
transition behaviour of an appropriately defined order parameter
was approximated to good accuracy in large networks of almost
arbitrary structure. In particular, the following expression for
the critical coupling strength was derived:
\begin{eqnarray}
k_c & = & \frac{k_0}{\lambda}.
\end{eqnarray}
Here $k_0$ is a constant, depending on the distribution of the
oscillators' intrinsic frequencies, and $\lambda$ is the spectral
radius of the network's adjacency matrix. Note that this estimate
requires full knowledge of the adjacency matrix. A less
restrictive estimate was obtained by introducing the additional
assumption that the components of the eigenvector associated with
the spectral radius are proportional to the vector of node
degrees. The expression thus obtained reads
\begin{eqnarray}
\label{eq:SyncThre} k_c & = & k_0\frac{\langle k \rangle}{\langle
k^2 \rangle},
\end{eqnarray}
which coincides with the result reported in \cite{ichinomiya04}. A
detailed account of the validity of the various assumptions
involved can be found in the paper \cite{restrepo05}. In the above
expression~(\ref{eq:SyncThre}), $\langle k \rangle$ and $\langle
k^2 \rangle$ denote the first and second moments of the node
degree distribution, respectively. As pointed out
in~\cite{ichinomiya04}, for scale-free networks with a power law
coefficient between $2$ and $3$, the second moment grows without
bound as the number of nodes tends to infinity. This would suggest
that, in such networks, there is no critical coupling in the
thermodynamic limit; or indeed no threshold for coherent
oscillations. This has been demonstrated not to be the case for
finite networks \cite{morenovega04,ichinomiya04}. Indeed, in
\cite{ichinomiya04} it is reported that there exists a clear
dependence between the critical coupling strength and the network
size. We draw attention to the fact that related observations have
been reported in the literature on disease propagation.
Particularly, the absence of an epidemic threshold has been
established as a characteristic feature of disease spread models
on (infinite) scale-free networks. Finite-size effects have also
been discussed in this context \cite{MayLlo01}. The similarity
between the physics of coupled oscillators and models of disease
spread has been discussed previously in \cite{ichinomiya04}. We
shall have more to say about this connection in the next section.

\underline{\bf Factors that Promote Synchronization}

Let us consider what structural properties of a network promote
synchronization. A recent study \cite{mcgraw05} suggests that one
factor might be the amount of \emph{clustering}. Indeed, the study
indicates that networks (Poisson or scale-free) that share the
same number of nodes, the same number of edges and the same degree
distribution, but have a different average clustering coefficient,
may have very different synchronization properties. In particular,
it was found that increasing the clustering coefficient of a
Poisson network leads to a more gradual transition from
incoherence to coherence. For scale-free networks, the effect was
more ambiguous in that increased clustering appeared to promote
the onset of synchronization at low coupling strengths,
suppressing the same at high coupling strengths. For moderate
coupling strengths the network would seem to split into several
dynamic clusters oscillating at different frequencies. The authors
proposed that scale-free networks with high clustering undergo two
separate transitions: a first transition to a partially
synchronized state, corresponding to the formation of clusters
oscillating at distinct frequencies; followed by a second
transition to full synchronization when the clusters are tuned to
a common frequency.

Other factors, reported in \cite{hong04}, in a study of
Watts-Strogatz small-world networks, include large maximum degree,
short characteristic path length, heterogeneity of the degree
distribution and a low value for the average betweenness
centrality. Among these factors, betweenness centrality was found
to account for the strongest correlations. Some of these findings
have been shown not to hold for other types of networks. Notably,
for scale-free networks, it appears that homogeneity in the degree
distribution, rather than heterogeneity would promote
synchronization \cite{nishikawa03}. This would contradict the
popular belief that because the average path length in
heterogeneous networks tends to be smaller than, for example, in
lattices, communication between oscillators would be more
efficient, which would amount to better synchronizability. Using
the ratio between the smallest (nonzero) and the largest
eigenvalue of the Laplacian as a measure of synchrony (which was
also the measure used in \cite{hong04}), the authors demonstrated
the opposite, namely that heterogeneity in a scale-free network
tends to inhibit rather than promote its ability to synchronize.

In~\cite{morenovega04}, it was demonstrated numerically that
(finite-size) scale-free networks of Kuramoto oscillators exhibit
a phase transition at a coupling strength that is inversely
proportional to the average node degree. In the same study, the
authors also investigated the `fitness for synchronization' of
particular network motifs, defining fitness as the (normalized)
coupling strength at which the probability that a motif
synchronizes first exceeds one half. The results suggested that
motifs with high interconnectedness are more prone to synchronize.
Interestingly, this ability to synchronize was found to be
correlated with the motif's natural conservation rate in the yeast
protein interaction network (see Section 5.3).

In a study involving $d$-dimensional lattices of coupled
oscillators \cite{hong05}, it was investigated what the minimal
dimension $d^*$ of a lattice should be in order for the
oscillators to synchronize in the limit of strong coupling. Based
on extensive simulations the authors conclude that $d^*=3$ for
frequency synchronization, and $d^*=5$ for phase synchronization.

For networks of Erd\"{o}s-R\'{e}nyi type, the authors
of~\cite{gong05} derived a lower bound on the critical average
degree, that is the smallest average degree for which
synchronization is possible. Moreover, if $p$ is the probability
that an edge is placed between a given pair of nodes in an ER
network, it was shown that networks with different values of $p$
share the same critical coupling strength, which is that of the
globally coupled network.

In small-world networks, the onset of phase and frequency
synchronization appears to depend strongly on the rewiring
probability when this probability is small, and no synchronization
whatsoever is observed when this probability is identically zero
\cite{hong02}. Interestingly, the synchronization behaviour
appears to be roughly the same for larger values of the rewiring
probability, suggesting some form of saturation to set in.

\subsection{Synchronization in the brain}
Having described various theoretical aspects of synchronization,
we shall close this section with a discussion on the proposed role
of synchronization in the brain.

\underline{\bf Synchronization and the Problem of Integration of
Information}

Synchronization has been put forward by some as the mechanism that
would make possible the integration of distributed neural activity
in our brain~\cite{varela01}. Others have argued against this.
Here we shall focus on the supporting evidence. What is neural
integration and what part does synchronization have to play in it?
Recent studies suggest that during processing of visual and
auditory stimuli, activities of functionally specific brain
regions are temporally aligned so as to produce a unified
cognitive moment. This would imply that an inability to
synchronize, due to abnormalities in the neural circuit for
instance, could have severe behavioral implications
\cite{odonnell02,spencer03}. An understanding of the mechanics of
this phenomenon may thus hold the key to devising new treatments
for neurological disorders.

It has been known for a long time that groups of neurons
\emph{within a single sensory modality} such as the visual cortex,
selectively synchronize their activities, supposedly to integrate
the particular features for which they encode. However, the fact
that this same kind of integration would take place \emph{across
different sensory modalities} was discovered only recently. In a
study reported by Roelfsema \emph{et al}.~\cite{roelfsema97}, five
cats were conditioned to press and release a lever in response to
particular visual stimuli. Electrodes were implanted at different
locations in the motor and visual cortices to monitor the
electrical activity during execution of the task. Coupling between
these brain areas was investigated using cross-correlation
analysis on pairs of LFP (Local Field Potential) traces. Tighter
coupling was observed when the animals were engaged in the
specific visuomotor task than when engaged in feeding or at rest.
Based on these and other findings, Varela \emph{et al}. have
suggested that ``large-scale synchrony is the underlying basis for
active attentive behaviour''. \cite{varela01,rodriguez99}.

In a more recent study~\cite{seidenbecher03}, it was investigated
how the interactions between selected areas in the hippocampus and
amygdala in fear-conditioned mice compare against those in
controls. The response of the fear-conditioned group indicated a
selective synchronization in the theta frequency range (4-7 Hz)
upon presentation of the conditioned stimulus, which was not found
in the control group. No significant synchronization was observed
in either group during presentation of the unconditioned stimulus.
It was argued that these results are indicative of a functional
relationship between theta rhythm synchronization and the
retrieval and expression of fear.


\underline{\bf Abnormal Neural Synchrony and Schizophrenia}

Assuming synchronization is the mechanism that underlies neural
integration, it seems reasonable to suppose that disruptions in
neural synchrony would impact one's behaviour. Interestingly, an
impaired ability to integrate information has long been identified
as one of the symptoms of Schizophrenia, which makes this disorder
particularly relevant in this context. Schizophrenia, a complex
and debilitating disease, is generally defined in terms of its
symptoms, which may be divided into (a) positive symptoms, which
include delusions, hallucinations, and incoherent thoughts; and
(b) negative symptoms, which include social withdrawal, poor
motivation, and apathy \cite{sawa02,kandel00}. In recent years, it
has been proposed that these cognitive and affective impairments
may be related to a defect in the mechanism believed to be
responsible for the integration of distributed neural activity,
that is, to gamma band synchronization
\cite{spencer03,odonnell02,hong04}.

A recent report supports this \cite{spencer03}: when a set of
Gestalt images were presented to a group of patients diagnosed
with Schizophrenia (SZ) and a group of Normal Control (NC)
subjects, a significant difference in neural orchestration between
the two was observed. A phase-locking response, persistent among
individuals from the NC group, but absent in the SZ group, was
hypothesized to reflect a feature-binding mechanism in the visual
cortex which would explain the more efficient task performance by
healthy individuals.

Further evidence for abnormal neural synchrony in Schizophrenia
was reported in~\cite{ioannides04}. In this study, two groups,
patients and controls, were presented with a set of images
depicting six basic human emotions, which they were to recognize.
The response of each individual was measured using whole head MEG
(Magnetoencephalogram). Local activity was averaged over a so
called region of interest (ROI) and a coherence score was computed
as the mutual information (MI) \cite{CoTh91} between ROIs. The MI
analysis revealed a very organized pattern of linkages for normal
subjects, as opposed to the overall disturbed linkages for
Schizophrenia patients. At some level, these results agree
 with the outcome of another
study~\cite{hong04}, which involved first-degree relatives of
patients with Schizophrenia. Gamma-band synchronization was found
to be reduced in first-degree relatives with Schizophrenia
Spectrum Personality Problems.

\underline{\bf A Theory of Neural Synchronization?}

It has been established beyond doubt that the processing of
particular audiovisual stimuli coincides with the temporal
synchronization of neural activities in functionally specialized
brain regions. In addition there is some evidence that patients
with Schizophrenia or related neurological disorders are more
likely to display abnormal patterns of synchrony than controls.
Meanwhile, the mechanics of this synchronization and its supposed
role in the integration of information remain poorly understood.
Most experimental studies resort to elementary statistical
techniques to conclude with confidence that some form of
synchronization takes place. Beyond that, there appears to be a
shortage of quantitative models; models that do not just extract
information from the data, but indeed attempt to explain the data.
With no disrespect for the seminal importance of Kuramoto's work,
and that of others' who have contributed to the theory of coupled
oscillators, it appears that we are still far removed from
effectively applying this theory in the context of the neural
synchronization problem. Fortunately, there is reason to believe
that this gap is closing fast, considering on the one hand the
rate at which measurement techniques are being refined, and, on
the other hand, some of the pioneering work that is being done on
the theoretical front.

\subsection{Summarizing Comments}

In this section we discussed some aspects of synchronization,
using the Kuramoto model of coupled oscillators as a starting
point. We reviewed recent results on the relation between network
structure and synchronizability.
\begin{itemize}
\item[(i)] the onset of synchronization in complex networks is
determined by a few key factors that include:~the average
clustering coefficient, the second moment of the degree
distribution, the maximum degree, the characteristic path length
and the average betweenness centrality. These factors may impact
different networks in different ways.

\item[(ii)] For scale-free network of infinite size and with a
power law exponent between 2 and 3, the value for the critical
coupling is zero. For finite-size scale-free networks, the
critical coupling is nonzero. We pointed out parallel results in
the disease propagation literature.

\item[(iii)] We discussed the role of synchronization in the brain
and argued that, ongoing efforts notwithstanding, there is a lot
of work to be done in the way of tuning the abstract mathematical
models of coupled oscillators to the experimentalist's needs.
\end{itemize}

\section{Network Structure and Disease Propagation}
\label{sec:Disease}

The final major topic that we shall consider here is the impact of
network structure on disease propagation models.  Given that
several of the novel network properties considered in the recent
past have been observed in social networks and in networks of
human sexual contacts \cite{Liletal01}, it is natural to ask what
effect these properties have on the spread of disease through such
networks. Given the emergence of new virulent diseases such as the
SARS virus and the Asian bird flu, the importance of understanding
the interaction between network structure and the dynamics of
disease propagation cannot be over-emphasised.  The current
section is organised as follows.  First, we shall discuss recent
numerical and theoretical work on the effect of different degree
distributions on the behaviour of classical epidemic models, with
particular emphasis on the effect of power-law distributions on
the so-called {\it epidemic threshold}.  We shall then discuss
extensions of this basic line of research which have attempted to
take into account finite-size effects correlations between the
degrees of connected nodes.  Finally, we shall discuss a number of
other issues pertaining to disease spread on networks, including
the containment of epidemics on different network topologies and
the evolution of different disease strains.

\subsection{Scale-free Networks and Epidemic Thresholds}
The mathematical theory of epidemics has been the subject of
intensive research for some time now and several different models
for disease spread have been developed.  A detailed discussion of
the properties of all of these models is well beyond the scope of
the current document, and the interested reader should consult
\cite{AndMay91, Het00}.  Here, we shall confine our discussion to
results concerned with two basic models of disease spread: the
{\it Susceptible-Infected-Susceptible} or {\it SIS} model and the
{\it Susceptible-Infected-Removed} or {\it SIR} model.  Much of
the recent work on disease propagation through networks has
focussed on these two core models.

In the SIS model, a population is divided into two groups: the
first (S) consists of susceptible individuals, who are not
infected but can contract the disease from members of the second
group (I) of infected individuals.  After a period of time, an
infected person recovers and then becomes susceptible again. Hence
no immunity is conferred  by contracting the disease and the
recovered infective can become infected again at a later time.  In
contrast, in the SIR model, a recovered infective is regarded as
being immune to the disease and cannot subsequently become
infected again.  Hence, the population is divided into three
groups in such models: susceptibles (S), infectives (I) and
removed or recovered (R).

There are two fundamental parameters associated with any SIS or
SIR model: the probability $\lambda$ of an infective passing on
the disease to a susceptible with whom they are in contact during
the period in which they are infective, and the rate $\nu$ at
which an infective recovers. In basic models of population
epidemiology, it is assumed that the population is homogeneously
mixed.  This essentially amounts to assuming that each individual,
or node, in the population has the same number of contacts.  Under
the assumptions of homogeneous mixing and a fixed population size,
the standard equations for the SIR model are given by
\cite{Mur02a, BraCas00}
\begin{eqnarray}
\label{eq:SIR} \frac{dS}{dt} &=& -\lambda SI \\
\nonumber \frac{dI}{dt} &=& \lambda SI - \nu I \\
\nonumber \frac{dR}{dt} &=& \nu I.
\end{eqnarray}
Here, the variables $S(t), I(t), R(t)$ represent the total number
of individuals in the susceptible, infected and recovered classes
respectively at time $t$.  From a network point of view, we can
consider the population as a graph, $G$, in which each individual
is represented by a node and each edge represents a contact or
connection between individuals, through which the disease can
spread.  In a homogeneously mixed population, each node $v$ in $G$
has the same degree, which would be equal to the mean degree,
$\langle k \rangle$, of the network. This assumption is only
reasonable for networks whose degree distributions are narrow,
meaning that the coefficient of variation, $C_V = \sqrt{\langle
k^2 \rangle / \langle k \rangle ^2 - 1}$, is very small.

Under the assumption of homogeneous mixing, the quantity $\iota_0
= \langle k \rangle \lambda / \nu $, represents the average number
of secondary infections that would result from the introduction of
a single infected individual into an entirely susceptible
population.  In this case, the introduction of an infective into
the population will result in an epidemic if the basic
reproductive number $R_0 = \iota_0$ is greater than one, while if
$R_0 < 1$, the disease will die out.  Thus, defining $\lambda_c =
\nu / \langle k \rangle$, an epidemic occurs if the spreading
rate, $\lambda$ satisfies $\lambda
> \lambda_c$ while the disease dies out if $\lambda < \lambda_c$.
The constant $\lambda_c$ is usually referred to as the epidemic
threshold.

While the assumption of homogeneous mixing might be reasonable for
the classical ER random graph models, it is entirely inappropriate
for BA and other scale-free networks with broad-tailed degree
distributions. In \cite{MayLlo01}, the dynamics of the SIR model
on heterogeneous networks of this type were studied.  It was
pointed out that for such networks, the basic reproductive number
$R_0$ is given by the formula
\begin{eqnarray}
\label{eq:HetR0} R_0 = \rho_0 (1 + C_V^2).
\end{eqnarray}
Now, in the limit as network size tends to infinity, for a
scale-free network with degree distribution of the form $P(k) \sim
k^{- \gamma}$ with $2 < \gamma < 3$, the coefficient of variation
$C_V$ of its node-degrees is infinite (more precisely, the second
moment $\langle k^2 \rangle$ diverges as the network size, $n$,
tends to infinity, while $\langle k \rangle$ remains finite).
Thus, for any non-zero spreading rate $\lambda$, the introduction
of an infective into the population can result in an epidemic.
Similar findings were also reported in \cite{PasVes01} for the SIS
model. These results lead to the somewhat surprising conclusion
that for scale-free networks, the epidemic threshold is
effectively zero.  This also follows from the following formula
for the epidemic threshold for scale-free networks with degree
distribution $P(k) \sim k^{-3}$, which was presented in
\cite{PasVes02b} (as well as a number of other sources).
\begin{eqnarray}
\label{eq:lamc} \lambda_c = \frac{\langle k \rangle}{\langle k^2
\rangle}
\end{eqnarray}
Note that this same formula has appeared above in the context of
coherent synchronization on random networks (\ref{eq:SyncThre}).

The authors of \cite{MayLlo01} also derived approximate
expressions for the fraction of nodes, $I$, in a scale-free
network that are ever infected for an SIR model of disease spread.
($I$ is usually referred to as the final epidemic size.)  First of
all, for scale-free networks with power-law exponent $\gamma = 3$,
they demonstrated that, essentially, $I$ decreases with decreasing
$\lambda$ as $Ce^{- (A/\lambda)}$ for constants $A, C$.  Using
approximate, mean-field arguments and simulations, a similar
result was derived in \cite{PasVes01} for the steady-state
prevalence of an SIS epidemic\footnote{The steady-state prevalence
is the fraction of infected nodes in the steady state.}. The
dependence of $I$ on $\lambda$ for networks with $2 < \gamma < 3$
was also calculated in \cite{MayLlo01} and it was established that
in this case, $I$ decays with decreasing $\lambda$ according to a
power law of the form $C (\lambda)^{1/ (3 - \gamma)}$.  In the
same paper, it was also shown that, for networks with $\gamma =
3$, the number of infected nodes of low-degree is small, while
many (essentially all) nodes of high-degree are infected.  These
findings are in agreement with those described in
\cite{Baretal05}, which indicate that disease spreads in a
hierarchical cascade from hub nodes to nodes with intermediate
degree to nodes with low degree.  These observations clearly have
significant implications for the development of containment
strategies. Specifically, they suggest that an effective
containment strategy would first and foremost target the hubs of a
network.  Similar recommendations have been made in
\cite{DezBar02a}.

Before we proceed, it should be noted that the results discussed
in the previous paragraph are based on a number of assumptions.
\begin{itemize}
\item[(i)]  The results described above were derived for the
limiting case of an infinite network or population, and rely on a
continuous approximation of the node-degree variable $k$.  It has
been noted in \cite{MayLlo01} that when finite size effects are
taken into account the epidemic threshold does not vanish but in
fact takes a positive value. \item[(ii)]   These results apply to
networks in which there is no correlation between the degrees of
connected nodes.  \item[(iii)] Finally, as with biological
interaction networks, the inferred scale-free nature of social and
sexual networks typically relies on sampled network data.   Hence,
in order to reliably apply the results discussed here, it is vital
to understand the effect of sampling on the identification of a
network's topology.
\end{itemize}
Later in this section, we shall describe the results of a number
of authors who have attempted to address some of these
limitations.

\subsection{Impact of Finite Size and Local Structure on Disease Spread}

The above results on the properties of SIS and SIR models on
scale-free networks were derived for the limiting case of networks
of infinite size.  Of course, real networks of social and sexual
contacts are finite and, for this reason, a number of authors have
studied the dynamics of disease spread on scale-free networks with
finitely many nodes. In \cite{PasVes02a}, the epidemic threshold,
$\lambda_c$, and the steady-state prevalence, $\rho$, for the SIS
model on finite scale-free networks were investigated.  It was
found that $\lambda_c$ is non-vanishing in this case, and formulae
approximating the dependence of $\lambda_c$ and $\rho$ on the
network size, $n$, were also derived. Note that while the epidemic
threshold is non-vanishing for the finite scale-free networks
studied in \cite{PasVes02a}, it is considerably smaller than for a
corresponding homogeneous network with the same average degree. In
fact for scale-free networks of size larger than 1000, the
threshold is at least one order of magnitude smaller than in the
homogeneous case.  These findings are largely in agreement with
the remarks on finite-size effects for SIR models made towards the
end of the paper \cite{MayLlo01}.  Note also the findings reported
in \cite{Hwaetal05} where the behaviour of the SIS model on two
different types of network with scale-free degree distributions
was studied numerically.  For both network types, the epidemic
threshold $\lambda_c$ is non-zero.  However, the dependence of
$\lambda_c$ on network size and the effect of the spreading rate
$\lambda$ on $\rho$ varied significantly between the two classes
of network, even for networks with the same underlying degree
distribution. These results demonstrate that it is possible for
two networks with the same degree distribution, but different
local structures, to exhibit significantly different behaviours
with respect to disease propagation.

The final observation in the previous paragraphs has motivated a
number of authors to study classes of scale-free networks in which
the degrees of neighbouring nodes are correlated.  Such networks
offer a more realistic picture of real social networks in which
such correlation is common.  In \cite{EguKle02} the SIS model was
studied on a class of highly-clustered scale-free networks.
Numerical simulations indicated that the highly clustered networks
behave in a qualitatively different manner than the usual
scale-free models, both with respect to the dependence of
steady-state prevalence $\rho$ on spreading rate $\lambda$ and to
survival probability of the disease.  Moreover, the authors of
this paper argue that for this highly structured class of
scale-free networks, there is a non-vanishing epidemic threshold
even in the limit as the network size, $n$, tends to infinity.
They further conjectured that the value of the threshold depends
on the degree correlations within the network rather than on the
degree distribution itself.

The relationship between the epidemic threshold and the degree
correlations in a scale-free network has been further investigated
in \cite{BogPas02, BogPasVes03}.  In \cite{BogPas02} the value of
the epidemic threshold is related to the largest eigenvalue of the
so-called connectivity matrix $C$, where $C_{kk'} = kP(k' | k)$.
Here $P(k' | k)$ represents the probability that a given link
emanating from a node of degree $k$ connects to a node of degree
$k'$.  For networks with no higher order correlations, they
demonstrate that the epidemic threshold is equal to the reciprocal
of the largest eigenvalue of $C$. Based on these results, in
\cite{BogPasVes03} conditions for the absence of an epidemic
threshold in scale-free networks with arbitrary two-point degree
correlation functions $P(k' | k)$ and degree exponents in the
range $2 < \gamma \leq 3$ were investigated. The principal result
of this paper established that in this case, provided the network
possesses no additional, higher order, structure, the epidemic
threshold is again zero in the limit of infinite network size.  We
should also note here the work described in \cite{VolVolBla02,
MorGomPac03} which further investigated the effects of degree
correlations and local structure on the dynamics of disease spread
in scale-free networks.

\subsection{Containment Strategies on Heterogeneous Networks}

One of the most fundamental issues in epidemiology is how to
design effective strategies for containing the outbreak of an
infectious disease.  One simple strategy would be mass
vaccination, in which (almost) every individual in the population
is vaccinated against a disease, and hence immune to it.  While
this can be an effective strategy for containing infectious
diseases, it is crude and operationally expensive. As a result,
there is great interest in alternative strategies which, although
perhaps slightly less effective, are much more economical in terms
of resources and logistics.  Recently, in \cite{DezBar02a,
PasVes02b}, the implications of power law degree distributions for
the design of immunization programmes was investigated using
mean-field approximations and numerical simulations.  The first
strategy considered was that of uniform random vaccination in
which individuals are uniformly selected at random and vaccinated.
However, while this strategy can work for homogeneous populations,
it is known to be ineffective in the heterogeneous case
\cite{AndMay91}. The findings in \cite{DezBar02a, PasVes02b}
suggest that for scale-free networks, and the SIS model of disease
spread, considerable improvements over uniform vaccination can be
achieved through targeting hub nodes within a network.  In fact,
two different approaches of this kind were suggested.  In the
first of these, nodes are vaccinated with probability proportional
to their degree, so that a greater proportion of nodes of high
degree are vaccinated than is the case for nodes of low degree.
The second strategy aims to specifically target hub nodes by
vaccinating all nodes in the network of degree higher than some
threshold $k_c$. While this appears to be more cost effective, in
terms of how many individuals need to be immunized in order to
eventually eradicate the disease, it relies on a fairly complete
knowledge of the network's topology.  As mentioned before in the
text, it is unrealistic to assume that we will have exact
knowledge of each individual's degree within the network and the
impact of sampling errors and inaccurate network data on
vaccination schemes needs to be analysed more thoroughly.

A disease containment strategy, aimed at controlling outbreaks of
smallpox was recently proposed in \cite{Eubetal04}. In this paper,
the social networks through which disease spreads were modelled as
bi-partite graphs \cite{Die00}. Such a graph has two distinct
types of vertices, which correspond to locations and individuals
respectively.  The results of this paper suggest a containment
strategy of targeted vaccination combined with early detection.
Early detection could be accomplished by placing sensors at
locations with high degree, that is, locations visited by many
people, while efficient vaccination is effected by targeting
long-distance travellers.  The impact of factors such as targeted
or mass vaccination schemes, withdrawing infected individuals to
their homes, and delays in introducing containment measures, on
the number of deaths caused by a smallpox outbreak was
investigated by numerical simulation. The results suggested that
the most significant factor was the early removal of infected
individuals to their homes with the next most influential factor
being the length of delay in implementing vaccination schemes.

In \cite{Becetal05}, motivated by the recent emergence of the SARS
virus, several intervention strategies for epidemic containment
were considered, and the impact of each strategy on the effective
reproduction number was determined.  In general, the results of
the paper suggest that combining different strategies is a good
idea, while the strategy of tracing and quarantining the contacts
of diagnosed cases was found to be particularly effective.  The
model studied in this paper incorporated several realistic aspects
of social structure.  For instance, given that people tend to be
more frequently in contact with individuals within their own
household than with people from other households, a distinction
was drawn between {\it within-household }transmission and {\it
between-household }transmission.  Furthermore, school-children and
the rest of the population were considered separately.  While the
manner of counting secondary infections, and the reproduction
number, used in this paper were somewhat non-standard, they have
the advantage of being analytically tractable and, moreover, the
number of ``offspring'' of a single infective, as counted in this
paper, is independent of the size of the infective's own
household.  Parameter values pertaining to the distribution of
household sizes were selected in accordance with given census
data.  Various control strategies were considered, including
exposure avoidance, isolating cases at diagnosis, closing schools,
quarantining affected household, and contact tracing. Apart from
the efficacy of the above mentioned strategy of tracing and
quarantining contacts of diagnosed cases, the results indicate
that if an emerging infection were to enter a juvenile population,
closing schools can reduce transmission significantly.

\subsection{Other Network Models and the General Theory of Disease
Spread on Networks}

In addition to the work discussed above on epidemic dynamics on
scale-free network, a number of authors have considered the
problem of disease spread on other network topologies.  For
instance, in \cite{SarKas05} the impact of dynamically adding
long-range links to regular one-dimensional lattices on the spread
of disease was studied.  Using the SIR model for disease spread,
they have shown that the resulting small-world network
\cite{WatStr98} structure exhibits a shortcut-dependent epidemic
threshold. An approximate expression for this threshold in terms
of the effective spreading rate and the effective recovery rate
was shown to be accurate over a large range of parameter values.
The authors also acknowledged the fact, previously stated
elsewhere \cite{Meyetal05, MayLlo01}, that the basic reproduction
number has limited use outside the homogeneous mixing paradigm.
They argue that this is particularly true for small-world networks
because ``the effect of a secondary infection caused by
nearest-neighbor transmission is different from the one caused by
a long-range jump'' \cite{SarKas05}.  Assuming a spreading
probability of one, so that susceptibles in direct contact with
infectives will become infected during the next iteration step, it
was shown that the epidemic saturation time, i.e. the time it
takes for 95\% of the susceptible population to become infected,
scales with $-log(n_0)$, where $n_0$ is the fraction of nodes
initially infected.  The scenario of spreading with near certainty
would correspond to the onset of an epidemic, and is used by the
authors to predict the final epidemic size as well the development
of an epidemic from its beginning stages.  The dynamics of the SIR
model and the related {\it susceptible-exposed-infected-removed
}(SEIR) model on small-world network were also investigated in the
paper \cite{Veretal05}.

Recently, in \cite{New02a} analytical techniques were developed
which can be used to derive exact solutions for a large class of
standard epidemiological models on a variety of networks. These
techniques are based on generating functions and allow for great
flexibility in terms of assumptions on network structure and
degree correlations. Further they can accommodate heterogeneity in
transmission rate and infectious period and allow for correlations
between parameters such as transmission rate and node degree. The
results derived in this paper include formulae for the epidemic
threshold and average outbreak size for the network classes
considered.  More recently, the problem of epidemic spread on
random graph models has been studied in a mathematically rigorous
fashion within the framework of Markov processes in
\cite{GanMasTow05}.  Here, the dependence of the final epidemic
size and the lifetime of an outbreak on graph parameters such as
the spectral radius of the network's adjacency matrix and the
isoperimetric number of the network was investigated. Some general
theorems as well as results for a variety of graph models
including the ER and scale-free models were derived for the SIS
and SIR models of disease spread.

The techniques developed in \cite{New02a} were then applied in
\cite{Meyetal05} in an effort to explain some puzzling aspects of
the recent SARS outbreaks. Specifically, the question of why these
outbreaks never led to an epidemic, given the relatively high
estimates for the basic reproduction number, was considered. Using
purely analytical tools, the authors derive expressions for the
likelihood that a small outbreak results in an epidemic in,
respectively, an urban network, a power-law network, and a Poisson
network. It turns out that (and this is confirmed by numerical
simulations) ``outbreaks are consistently less likely to reach
epidemic proportions in the power-law network than in the
others''.  It should also be noted that it was shown that for all
networks there is a nonzero probability that an outbreak does not
become an epidemic, even when the spreading rate of a disease
exceeds the epidemic threshold. By contrast, in the paradigm of
homogeneous mixing, an epidemic will occur with certainty whenever
the basic reproduction number is greater than unity. Other
interesting findings are that:
\begin{itemize}
\item[(i)] The likelihood of an outbreak is a monotonically
increasing function of the degree of the first infective;
\item[(ii)] When the transmissibility of disease is far above the
epidemic threshold, the risk of an epidemic is very high even for
small initial outbreaks, at least in the case of urban networks.
\end{itemize}

Finally, we note that the evolution of diseases on local and
global networks has been studied in \cite{ReaKee03}. The basic
premise of this work was that different disease strains adapt to
compete for resources (susceptible hosts). In the model proposed
here, adaptation corresponds to a random mutation of both the
transmission rate and the infectious period, which takes place
whenever a new infection occurs. As the authors point out, in
mean-field models this type of evolution would result in runaway
behavior with selection for ever higher transmission rates and
ever longer infectious periods. By contrast, both spatial
heterogeneity in local networks and the presence of shortcuts in
global networks appear to constrain the evolutionary dynamics, to
the effect that the rate of adaptation is generally slower (in the
case of a global network, the transmission rate even saturates at
some finite value) and the variability (in the dynamics) higher
than in mean-field models. Simulation results suggest that in
networks with many long-distance connections and a low clustering
coefficient, disease strains with conservative transmission rates
and long infectious periods are most likely to survive. By
comparison, for networks with strong local connectivity the
fittest strains are those that have high transmission rates and
relatively short infectious periods.

\subsection{Summarizing Comments}
\begin{itemize}
\item[(i)]  The structure of a social network can have a
significant impact on the dynamics of disease propagation.  In
particular, it has been shown for scale-free networks, in the
limiting case of infinitely many nodes, that the epidemic
threshold is zero.  This would mean that any non-zero spreading
rate could lead to an epidemic.  \item[(ii)] The previous fact was
initially established for uncorrelated scale-free networks of
infinite size.  For scale-free networks of finite size, the
epidemic threshold is non-vanishing but considerably smaller than
in the case of a homogeneously mixed population. \item[(iii)]
Results have recently been derived giving conditions under which
the epidemic threshold will be zero for scale-free networks with
degree correlations, in the limiting case of networks of infinite
size. \item[(iv)]  The dynamical behaviour of epidemics on
networks with heterogeneous degree distributions has implications
for the design of strategies for containing outbreaks.  In
particular, the targeting of nodes, or individuals, of high degree
can offer significant improvements over random immunization
programmes.
\end{itemize}

\section{Conclusions and Directions for Future Research}
\label{sec:Conc} The need for a more systematic approach to the
analysis of living organisms, alongside the availability of
unprecedented amounts of data, has led to a considerable growth of
activity in the theory and analysis of complex biological networks
in recent years.   Networks are ubiquitous in Biology, occurring
at all levels from biochemical reactions within the cell up to the
complex webs of social and sexual interactions that govern the
dynamics of disease spread through human populations.  Over the
last few years, several core themes and questions in biological
network analysis have arisen from pressing problems in Biology and
Medicine.  For instance, while the research on bio-molecular and
neurological networks is still at a relatively early stage, a
comprehensive understanding of these networks is needed to develop
more sophisticated and effective treatment strategies for diseases
such as Cancer and Schizophrenia.  Other aspects of this line of
research have been motivated by the need to determine the
biological role of unannotated genes or proteins.  On the other
hand, at the level of social networks, future approaches to
epidemic containment will need to take into account the interplay
between network topology and dynamics.

Our aim in this article has been to provide as comprehensive an
overview as possible of the uses of Graph Theory and Network
Analysis within Biology, and to point out problems in Graph Theory
that arise from the study of biological networks. Specifically, we
concentrated on the following five broad topics.
\begin{itemize}
\item[(i)] {\it Structural identification and modelling of
bio-molecular networks}

Recent advances in high-throughput techniques have led to the
construction of maps of protein-protein interaction,
transcriptional regulatory and metabolic networks for a variety of
organisms.  Numerical investigations of the properties of these
network maps, described in Section \ref{sec:models}, indicate that
they tend to have short characteristic path lengths, high
clustering coefficients and scale-free degree distributions.
Motivated by these observations, mathematical models such as the
Barabasi-Albert scale-free network and Duplication-Divergence
models have been proposed for protein interaction and genetic
networks.  However, the experimental techniques on which these
network maps are based are prone to high rates of false positive
errors, and typically only cover a fraction of the network's
nodes.  The development of more accurate and reliable experimental
methodologies is of course of vital importance for future research
on the structure of bio-molecular networks. On a more theoretical
level, two of the most significant issues that need to be
addressed in this area are the sampling properties of complex
networks and the impact of data inaccuracies on the identification
of network statistics such as the degree distribution.

\item[(ii)] {\it Centrality measures and essentiality in gene and
protein networks}

Much of the research on applying centrality measures to
bio-molecular networks has focussed on the prediction of gene or
protein essentiality.  In most of the studies discussed in Section
\ref{sec:Centrality}, the centrality score of a node was found to
be indicative of its likelihood to be essential.  In particular,
this appears to be true for degree centrality, betweenness
centrality and eigenvector centrality measures.  However, there is
no clear evidence that the more complex centrality measures
perform any better than degree centrality.  A major source of open
problems in this area is the robustness of centrality measures to
data inaccuracies.  Once again, this issue is very important for
the reliable application of these techniques to biological data.

\item[(iii)] {\it Motifs, modules and the hierarchical structure
of bio-molecular networks}

The research on motifs described in Section \ref{sec:motifs} has
helped to clarify the structural organisation of complex
biological networks.  Furthermore, the motifs of a network appear
to characterise it in some sense, and motifs such as the
feed-forward loop seem well suited to specific information
processing tasks.  However, while the motifs of a network
represent statistically significant patterns, their precise
biological significance and the mechanisms behind their emergence
are only partially understood.  To date much of the work on motifs
has been numerical in nature, and the theoretical analysis of the
motif profiles in mathematical network models is a potentially
rich source of open and challenging problems.  Moreover, analysis
of this type will provide further insights into how accurately
models such as Duplication-Divergence network describe real
biological networks.  A related area of research, discussed in
Section \ref{sec:motifs}, is the identification of functional
modules and the prediction of protein function based on network
topology.  The latter problem is of considerable importance and
the results discussed in the text indicate the potential of graph
theoretical approaches to this question.

\item[(iv)] {\it Synchronization, network topology and
neurological function}

Among the key issues in the study of complex networks is the
question as to how indicative a network's topology is of its
overall behaviour. Studies on systems of coupled oscillators
suggest that, as far as fitness for synchronization is concerned,
there are at least five structural properties that qualify as
important indicators. These are: average betweenness centrality,
average clustering coefficient, maximum degree, node degree
variance, and characteristic path length.

The dynamics of synchronization on scale-free networks are
characterized by a two-stage transition, initiated at low coupling
strengths by the formation of distinct clusters oscillating at
distinct frequencies, followed by a process of alignment at high
coupling, during which the different clusters are tuned to a
common frequency. A second characteristic property of scale-free
networks is that, in the limiting case of infinitely many nodes,
the threshold for the onset of synchronization vanishes.

At present, the majority of results on synchronization in complex
networks appear to have been obtained using a combination of
approximations and extensive simulations. As such, there is a
clear need for a rigorous mathematical analysis to support, and
underpin, the numerical findings. Also there is work to be done in
applying the abstract mathematical models of phase-coupled
oscillators to the analysis of experimental data.

\item[(v)] {\it Network structure and epidemic dynamics}

The interplay between epidemic dynamics and network structure is
vital for understanding and containing the spread of infectious
diseases.  The numerical studies and mean-field analyses discussed
in Section \ref{sec:Disease} have shown that a scale-free topology
can significantly reduce the epidemic threshold, making the
outbreak of epidemics more likely in networks with such a
structure.  Network topology also has an impact on the
effectiveness of immunization schemes for containing epidemic
outbreaks.  In particular, for networks with a scale-free
topology, the targeted immunization of nodes of high degree offers
substantial improvements over uniform random immunization.  Of
course, the reliable identification of social network structure is
vital for the practical implementation and interpretation of such
results.  One important direction for future research in this area
is the extension of recent results to incorporate the effects of
sampling and data noise on epidemic dynamics on networks and
containment strategies.
\end{itemize}
%

To finish, it is our hope that this article will be of assistance
to the broad community of researchers working on the study of
biological networks, by highlighting recent advances in the field,
as well as significant issues and problems that still need to be
addressed.

\subsubsection*{Acknowledgements}
This work was partially supported by Science Foundation Ireland
(SFI) grant 03/RP1/I382 and SFI grant 04/IN1/I478. Science
Foundation Ireland is not responsible for any use of data
appearing in this publication.

\bibliographystyle{plain}
\bibliography{Networks}
\end{document}